\documentclass[11pt,a4paper]{article}
\pdfoutput=1
\usepackage{diagbox}
\usepackage{jheppub}
\usepackage{graphicx}
\usepackage{amsmath}
\usepackage{amssymb}
\usepackage{amsthm}
\usepackage{mathtools}

\usepackage{mathrsfs}
\input epsf
\usepackage{epsfig}
\usepackage{tikz}
\usepackage{subfigure,tikz}
\usepackage{graphics}
\usepackage{graphicx}
\usepackage{diagbox}
\usepackage{mathtools}
\usepackage{xcolor}
\usepackage[thicklines]{cancel}

\newcommand{\bea}{\begin{eqnarray}}
	\newcommand{\eea}{\end{eqnarray}}
\newcommand{\bean}{\begin{eqnarray*}}
	\newcommand{\eean}{\end{eqnarray*}}
\newcommand{\nn}{\nonumber \\}

\def\newline{{\hspace{15pt}}}

\def\abs#1{\left| #1\right|}

\def\det{\mathop{\rm det}}

\def\eref#1{(\ref{#1})}

\def\d{{\rm d}}
\def\wt{\widetilde}

\def\a{{\alpha}}

\def\b{{\beta}}

\def\d{\partial}
\def\dd{\text{d}}

\def\eps{\epsilon}
\def\what{\widehat}
\newcommand{\overtr}[1]{(\overline{#1})}

\newcommand{\tensorc}[3]{C^{(#1)}_{{#2}\to {#2;\widehat{#3}}}}
\newcommand{\tensorselfc}[2]{C^{(#1)}_{{#2}\to {#2}}}

\newcommand{\measure}[2]{{\langle{#1}\mathrm{d}^{#2}{#1}\rangle}}

\newcommand{\lt}[2]{\text{Lt}_{#1}^{(#2)}}

\newcommand{\adtr}[1]{\dot{(\overline{#1})}}
\def\co{\,,}
\def\ed{\,.}
\def\label#1{\label{#1}%
	\smash{\hbox to0pt{\raise1ex\hbox{\tiny[#1]}\hss}}}

\newcommand{\parall}[2]{{#1}\ /\kern -0.8em / \  {#2}}

\title{\boldmath %Non-trivial Recursion Relation of One-loop Feynman Integral Reduction
Nontrivial One-loop Recursive Reduction Relation}

\author[a]{Tingfei Li}

\affiliation[a]{Zhejiang Institute of Modern Physics, Zhejiang University, Hangzhou, 310027, P. R. China }

\emailAdd{tfli@zju.edu.cn}

\abstract{In \cite{Feng:2022rwj}, we proposed a universal method to reduce one-loop integrals with both tensor structure and higher-power  propagators. But the method is quite redundant as  it does not utilize the results of lower rank cases when addressing certain tensor integrals. Recently, we found a remarkable recursion relation \cite{Feng:2022iuc,Feng:2022rfz}, where a tensor integral is reduced to lower-rank integrals and \textit{lower terms} corresponding to  integrals with one or more propagators being canceled. However, the expression of the lower terms is unknown.  In this paper, we derive this non-trivial recursion relation for non-degenerate and degenerate cases and provides an explicit expression for the lower terms, thus simplifying and speeding up the reduction process.   }
\keywords{One-loop, Recursion relation, Reduction}

\begin{document}
	\maketitle
	\flushbottom

	\section{Introduction}
	In recent years we have witnessed enormous progress in computing and understanding analytic structures of scattering amplitudes. At the one-loop level, it is well-known that a general one-loop integral in $D=(4-2\eps)$-dimensions can always be reduced to a linear combination of one-loop scalar integrals (known as \textit{master integrals}) with rational functions of external variables as their reduction coefficients~\cite{Brown:1952eu,Melrose:1965kb,Passarino:1978jh,tHooft:1978jhc,vanNeerven:1983vr,Bern:1993kr,Bern:1994cg,Campbell:1996zw,Devaraj_1998,Fleischer:1999hq,Binoth:1999sp,Duplancic:2003tv,Ellis:2007qk,Ossola_2007}. The master integrals at one-loop level, referred to as tadpoles, bubbles, triangles, boxes, and pentagons based on the number of their propagators, are well-known. Thus, the main problem of one-loop integrals is calculating the reduction coefficients. There are several methods available to deal with reduction at the integrand and integral levels, such as Integration-By-Parts (IBP)~\cite{Chetyrkin:1981qh,Tkachov:1981wb,Laporta:2000dsw,vonManteuffel:2012np,vonManteuffel:2014ixa,Maierhofer:2017gsa,Smirnov:2019qkx}, Passarino-Veltman (PV) reduction~\cite{Passarino:1978jh}, Ossola-Papadopoulos-Pittau(OPP) reduction~\cite{Ossola:2006us,Ossola:2007bb,Ellis:2007br}, Unitarity cut ~\cite{Bern:1994zx,Bern:1994cg,Bern:1997sc,Britto:2004nc,Britto:2005ha,Anastasiou:2006gt,
    Britto:2006sj,Anastasiou:2006jv,Britto:2006fc,Britto:2007tt,Britto:2010um}, Intersection number \cite{Mastrolia:2018uzb,Mizera:2019ose,Frellesvig:2019uqt,Frellesvig:2019kgj,Mizera:2019vvs,Frellesvig:2020qot,Caron-Huot:2021xqj} and etc. In this paper, we mainly discuss one-loop analytic tensorial reduction, which has a long history \cite{Stuart:1987tt,Devaraj_1998,vanOldenborgh:1989wn,Bern:1992em,DAVYDYCHEV1991107,Campbell:1996zw,Binoth:1999sp,Denner:2002ii,Denner:2005nn,Fleischer:2010sq}.

	Recently, the analytical structure of one-loop integrals is studied by investigating Feynman parametrization in the projective space for its compactness and the close relation to geometry~\cite{Arkani-Hamed:2017ahv,Gong:2022erh}. Inspired by these papers, we find it could be convenient to do reduction for one-loop integrals in projective space. Furthermore the symmetry and compactness of reduction coefficients are illustrated clearly by this method, as shown in  \cite{Feng:2022rwj}. However, with this technique, we have to expand a general one-loop integral into the combination of $E_{n,k}[V^i]$ first, then reduce every $E_{n,k}[V^i]$ to the basis, after that, we sum over all contributions to obtain the final reduction result. Here $E_{n,k}[V^i]$ denotes integrals defined in projective space.
	\begin{align}
		E_{n,k}[T]\equiv \int_{\Delta}{\measure{X}{n-1}T[X^k]\over (XQX)^{n+k\over 2}}\ ;~~ T[X^k]\equiv T_{I_1I_2\ldots I_k}X^{I_1}X^{I_2}\cdots X^{I_k}
	\end{align}
	where  $T$ is a general rank-$k$ tensor and $\Delta$ is a simplex in $n$-dimensional space defined by $X_I>0,\forall I=1,2,\ldots,n$.  The homogeneous coordinates $X_I$ are denoted by a square bracket $X=[x_1:x_2:\ldots:x_n]$, and two coordinates are  equivalent to each other up to a scaling, \textit{i.e}., $[x_1:x_2:\ldots:x_n]\sim [kx_1:kx_2:\ldots:kx_n]$ for any $k\neq 0$. The measure in the projective space is given by the differential form
	\begin{align}
		\measure{X}{n-1}={\epsilon^{I_1,I_2,\ldots,I_n}\over (n-1)!}X_{I_1} \dd X_{I_2}\wedge \dd X_{I_3} \wedge \ldots \wedge \dd X_{I_n}\ed
	\end{align}
	The matrix $Q$ appearing in the denominator $
	XQX=Q^{IJ}X_IX_J $ has component $Q^{IJ}=Q_{IJ}=\left(m_I^2+m_J^2-(q_I-q_J)^2\right)/2,I,J=1,\ldots, n$.
	For simplicity, we denote $E_{n,k}[V^i]\equiv E_{n,k}[T=\otimes V^i\otimes L^{k-i}]$ for arbitrary vector $V$ and the constant vector $L\equiv [1:1:\cdots:1]$,
	\begin{align}
		\otimes V^i\otimes L^{k-i}\equiv \underbrace{V\otimes V\otimes \dots V}_{i\ \text{times}} \otimes \underbrace{L\otimes L\otimes \dots L}_{k-i\ \text{times}}.
	\end{align}
		\begin{table}[h]
		\centering
		\renewcommand{\arraystretch}{1.5}
		\begin{tabular}{|>{\centering\arraybackslash}p{1.6cm}|>{\centering\arraybackslash}p{2.0cm}|>{\centering\arraybackslash}p{2.0cm}|>{\centering\arraybackslash}p{2.0cm}|>{\centering\arraybackslash}p{2.0cm}|}
			\hline
			\diagbox{$n$}{time$/s$}{$r$} & $1$ & $2$ & $3$ &$4$\\
			\hline
			$2$ & 0.0188& 0.101 & 0.349& 0.564  \\
			\hline
			$3$ & 0.0221 & 0.152 & 5.431 & 86.79\\ 
			\hline
			$4$ & 0.0271 & 0.427 & 432.1& \slash \\ 
			\hline
			$5$ & 0.0394 & 2.846 & 2307& \slash \\ 
			\hline
		\end{tabular} 
		
		\caption{Actual running time of tensor reduction by direct expansion. First, we expand a one-loop tensor integral according to \eref{UniversalCoeff}, where we set $v=n$ and $S=V$. Next, we choose $\widetilde{Q}=Q^{-1}$ in \eref{general-resur} to extract all $V$s from $T=\otimes V^i\otimes L^{k-i}$ iteratively. Finally, we reach a summation of scalar $E_{n,n-D'}$'s with the ``wrong" dimension $D'=D+2s$, where $s\in\mathbb{Z}$. Therefore, for every term, we need to shift dimension $D'$ to $D$. Finally, we sum over all contributions to obtain the reduction coefficients. We ran our code in \textsf{Mathematica} and found that the time cost grows sharply. For more details, please see the examples in Section 3 of \cite{Feng:2022rwj}.   }
		\label{running time}
	\end{table}
 The order of $V$ and $L$ does not matter, as both are contracted with $X$. While the compact recursion relation makes the reduction faster, we found that the running time increases sharply with the number of propagators $n$ and the rank $r$, as shown in Table \ref{running time}. However, using the recursion relation of $E_{n,k}[V^i]$ to investigate the analytical structure of reduction coefficients is not as convenient, as it provides only parts of the entire expression.
 Recently, we discovered a remarkable recursion relation for one-loop reduction, where a tensor integral is reduced to lower-rank integrals. The resulting \textit{lower terms} correspond to integrals where one or more propagators are canceled 
	\begin{align}
		I_{n}^{(r)}={1\over \abs{G}}\left[A_rI_{n}^{(r-1)}+B_rI_{n}^{(r-2)}+\text{Lower Terms}\right] \label{original-form}
	\end{align}
	where $\abs{G}$ is the Gram determinant of $I_{n}$ and the tensor integrals are defined as
	\begin{align} I^{(r)}_{n}\equiv\int {d^D \ell\over i\pi^{D/2}}  {(2R\cdot \ell)^r\over (\ell^2-M_0^2)\prod_{j=1}^{n} ((\ell-K_j)^2-M_j^2)}\ed~~~\label{old-one-loop}
	\end{align}
	The formula \eref{original-form} is proved by considering acting on it with two differential operators ${\cal D}_i\equiv K_i\cdot \d_R,{\cal T}\equiv \d_R\cdot \d_R$ \cite{Feng:2022rfz}. The expressions for $A_r,B_r$ are known while the $\textit{lower terms}$ is still a mystery.  In this paper, we derive the explicit form of the lower terms by applying the universal recursion relation of $E_{n,k}[V^i]$. We obtain two forms of the lower terms, the first one involves integrals in $(D-2)$-dimensional space while the second one involves integrals in the same $D$-dimensional space\footnote{While writing this paper, we became aware of similar results obtained by Chen et al. \cite{Chen:2022jux}, who used the IBP method in Baikov representation. Our work complements theirs by providing an alternative approach that emphasizes geometric interpretations and explicit constructions. }. The recursion relation not only makes the reduction process extremely simple and effective, but also serves as a tool to study the analytical structures of reduction coefficients like singularities.
	
	The paper is structured as follows: In
	Section \ref{review}, we revisit the reduction technique in projective space. In Section \ref{nonde-Q}, we give the recursion relation for non-degenerate $Q$ and then illustrate it with some examples. The proof is listed in the Appendix \ref{appdix:lowerterm}. In Section \ref{sec:deQ}, we discuss how to deal with the degenerate case and show our method with some examples. Finally, we provide  some discussion in Section \ref{sec:discussion}.

	\section{Review of one-loop reduction in projective space}
	\label{review}
	In recent work~\cite{Feng:2022rwj}, we study  general one-loop integrals in $D$-dimensional spacetime with both tensor structure and higher poles\footnote{Here to manifest permutation symmetry of propagators we use the notations differing from \eref{old-one-loop}. }
	\begin{align} I^{(r)}_{\mathbf{v}_n;D}\equiv\int {d^D \ell\over i\pi^{D/2}} {(2R\cdot \ell)^r\over \prod_{j=1}^n ((\ell-q_j)^2-m_j^2)^{v_j}}~~~~\label{TensorPoleR}
	\end{align}
	where the loop momentum $\ell$, auxiliary vector $R$ and external momentum $q_j$    live in $D=(d-2\eps)$-dimensional spacetime. 
    We denote $\mathbf{v}_n=\{v_1,v_2,\ldots,v_n\}$  the power list of the $n$ propagators. In the formula, we introduce  an  auxiliary vector $R$ to make the expression compact\footnote{ The introduction of the auxiliary vector $R$ can improve the efficiency of reduction as shown in \cite{Feng:2021enk, Hu:2021nia, Feng:2022uqp, Feng:2022rfz, Chen:2022jux,Feng:2022hyg} for one-loop integrals and \cite{Feng:2022iuc, Chen:2022lue} for higher loops.}. one can see any general tensor structure can be recovered by applying differential operators of $R$ on the standard expression, for example
	\bea
	\int {d^D \ell\over i\pi^{D/2}}  {\ell^2 \ell\cdot K\over \prod_{j=1}^n ((\ell-q_j)^2-m_j^2)^{v_j}}\propto (K\cdot \d_R)(\d_R\cdot \d_R)I^{(3)}_{\mathbf{v}_n;D}\ed
	\eea
After some algebra \cite{Feng:2022rwj}. 	we find \eref{TensorPoleR} can be written as a compact form in terms of $E_{n,k}$
	\bea
	I_{\mathbf{v}_n;D}^{(r)}&=&\sum_{i=0}^{r} {i!\Gamma(v-D/2-r)\over (-1)^{v+r}  (v-n+i)!}\mathscr{C}^{D/2+r-v}_{r,i}(R^2)^{r-i\over 2} E_{n,2v-n-D-r+i}[ S^{v-n+i}]\Big\vert_{t^iz^{\mathbf{v}_n-1}}~~~~~~~\label{UniversalCoeff}
	\eea
	where $v\equiv \sum_{i=1}^n v_i,S\equiv tV+Z,Z\equiv\sum_{i=1}^{v-n}z_iH_i$ and $\vert_{t^iz^{\mathbf{v}_n-1}}$ means to take the coefficient of $t^iz^{\mathbf{v}_n-1}\equiv t^i\prod_{i=1}^n z_i^{v_i-1}$. We have defined the vectors $V$, $H_i$ as %\footnote{We set $q_1=0$ throughout the paper.}
	\bea
	V\equiv [R\cdot q_1: R\cdot q_2:\ldots: R\cdot q_n],~~~~ H_i\equiv [0: \ldots 0:\mathop{1}\limits_{i-{\text{th}}}:0:\ldots: 0]\co
	\eea
	and the expansion coefficient in \eref{UniversalCoeff} is
	\bea
	\mathscr{C}^k_{r,i}=\frac{2^r r! \Gamma\big(  {r-i+1\over 2}\big)}{\sqrt{\pi }
		i! (r-i)!}\prod_{j=1}^{{r+i\over 2}}(k+1-j)=\frac{2^r r!k! \Gamma\big(  {r-i+1\over 2}\big)}{\sqrt{\pi }
		i! (r-i)!(k-{r+i\over 2})!}\co~~~ {r-i\over 2}\in \mathbb{N},
	\eea
	where we require $i$ to have the same parity as $r$.  We denote $I_{v;D}^{(r)}$ as the one-loop integrals before taking coefficient of $t^iz^{\mathbf{v}_n-1}$, whose reduction result is
	\begin{align}
		I_{v;n}^{(r)}=\sum_{j=0,\mathbf{b}_j}C^{(r)}_{v;n\to n;\what{\mathbf{b}_j}}I_{n;\what{\mathbf{b}_j}}
	\end{align}
	where $\mathbf{b}_j\equiv \{b_1,b_2,\ldots,b_j\}$ is the length-$j$ label list of the propagators being canceled. It is found that the reduction of general one-loop integrals is solved by a simple recursion relation for $E_{n,k}[T]$
	\bea
	E_{n,k}[(Q\widetilde{Q}T)]=\alpha_{n,k}\left[\sum_{b=1}^nE^{(b)}_{n-1,k-1}[(H_b\widetilde{Q}T)]+\sum_{k-1\ \text{ways}}E_{n,k-2}[\text{tr}_{\widetilde{Q}} T]\right]\co~~~~\label{general-resur}
	\eea
	where  $\alpha_{n,k}={1\over n+k-2}$ and $\wt{Q}$ is an arbitrary symmetric matrix. In the formula we need to sum the $(k-1)$ ways to contract indices between $\wt{Q}$ and $T$
	\begin{align}
		(Q\widetilde{Q}T)^{I_1I_2,\ldots,I_k}&=Q^{I_1J_1}\widetilde{Q}_{J_1J_2}T^{J_2,I_2,I_3,\ldots,I_k}\co\nn
		\sum_{(k-1)  \text{ways}}(\text{tr}_{\widetilde{Q}} T)^{I_3,\ldots,I_k}&=\widetilde{Q}_{I_1I_2}T^{I_1I_2I_3\ldots I_k}+\widetilde{Q}_{I_1I_2}T^{I_1I_3I_2 \ldots I_k}+\cdots +\widetilde{Q}_{I_1I_2}T^{I_1I_3\ldots I_{k}I_2}\ed
	\end{align}
	The $(n-1)$-dimensional integral $E^{(b)}_{n-1,k-1}$ means the integral obtained from the $n$-dimensional $E_{n,k}$ by deleting the $b^{\text{th}}$  component of $X$.
	The reduction process is achieved by following steps:
	\begin{itemize}
		\item Step 1: Expanding a general one-loop integral according to \eref{UniversalCoeff}.
		\item Step 2: Depending on whether $Q$ is degenerate, take $\widetilde{Q}=Q^{-1}$ or $\widetilde{Q}=Q^*$ in \eref{general-resur} to reduce every term in the expansion \eref{UniversalCoeff} to scalar basis.
		\item Step 3: Sum over all contributions.
	\end{itemize}
	In this paper, we mainly discuss pure tensor reduction which is involved in the real scattering process. Consider non-degenerate $Q$, we can choose $\widetilde{Q}=Q^{-1}$, and then \eref{general-resur} becomes
	\bea
	E_{n,k}[V^{i+1}]
	=\alpha_{n,k}\left [\overtr{H_bV}E^{(b)}_{n-1,k-1}[V^{i}]+{i }\overtr{VV}E_{n,k-2}[V^{i-1}]+(k-i-1)\overtr{VL}E_{n,k-2}[V^i]\right]\ed\nn ~~~\label{Enk-ViRecursion}
	\eea
	where we suppress the summation of $b=1,2,\ldots,n$ and  have defined the compact notation
	\bea
	\overtr{WZ}\equiv \sum_{I,J} W^I(Q^{-1})_{IJ}Z^J,~ \overtr{WZ}_{(\{j\})}\equiv \sum_{I,J\not\in \{j\}}W^I(Q_{(\{j\})}^{-1})_{IJ}Z^J\co
	\eea
	where $\{j\}$ is the sub-list of the label list of removed propagators and $Q_{(\{j\})}$ denotes of $Q$ matrix of corresponding one-loop integrals. Although we have used compact notations in every iteration to make all steps of the reduction process purely algebraic, it can be time-consuming for large $n$ and $r$. To estimate the running time, we can consider an even rank of $r=2m$ for simplicity. First, there are $r/2=m$ terms $E_{n,n-D-2m+i}[V^i]$ to deal with, where $i=0,2,4,\dots,2m$. We can regard $D'=n-k$ as the effective dimension for $E_{n,k}[V^i]$, and in the expansion \eref{UniversalCoeff}, every term has a "wrong" dimension of $D'=D+2m-i>D$. We also notice that the last two terms in \eref{Enk-ViRecursion} increase the dimension from $D'$ to $D'+2$. Each iteration of \eref{Enk-ViRecursion} extracts one $V$ or two $V$'s at the cost of higher dimensions. Roughly speaking, the time to reduce $E_{n,n-D-\delta D}[V^i]$ satisfies the recursion relation
	 \begin{align}
	 	T_{n}[\delta D,V^i]=nT_{n-1}[\delta D,V^{i-1}]+T_{n}[\delta D+2,V^{i-1}]+T_{n}[\delta D+2,V^{i-2}]\ed \label{time-recur}
	 \end{align} 
  By applying \eref{Enk-ViRecursion} iteratively, we can finally reach the scalar integrals with wrong dimension. The last step is to shift higher dimension back to $D$
  \bea
  E_{n,k}={E_{n,k+2}-\alpha_{n,k+2}\overtr{H_jL}E^{(j)}_{n-1,k+1}\over \b_{n,k+2}\overtr{LL}}\ed\label{Dimension-Lower}
  \eea
  So the time for dimension shifting satisfies 
  \begin{align}
  	T_{n}[\delta D,V^0]=nT_{n}[\delta D-2,V^0]+T_{n-1}[\delta D,V^0]\ed \label{time-dim-shift-recur}
  \end{align}
 The total time to reducing an one-loop tensor integral is
  \begin{align}
  	T_{n,r}=\sum_{i\le r,r-i=\text{even}}T_{n}[\delta D=r-i,V^i]\ed \label{total-time}
  \end{align}
  For simplicity, we ignore the time cost of summing over all terms in each step, and let $T_{n}[\delta D=0,V^0]=t_0$. We have listed the theoretical running times in Table \underline{\ref{running time-theory}}. However, when we compare it with Table \underline{\ref{running time}}, we observe that for $n=3,4$, the actual running time grows faster than what we have analyzed. This is because we have not accounted for the time cost of summing terms for every iteration in \textsf{Mathematica}. These summation processes can be very cumbersome for larger $n$ and $r$.
\begin{table}[h]
	\centering
	\renewcommand{\arraystretch}{1.5}
	\begin{tabular}{|>{\centering\arraybackslash}p{1.74cm}|>{\centering\arraybackslash}p{2.0cm}|>{\centering\arraybackslash}p{2.0cm}|>{\centering\arraybackslash}p{2.0cm}|>{\centering\arraybackslash}p{2.0cm}|>{\centering\arraybackslash}p{2.0cm}|}
		\hline
		\diagbox{$n$}{time$/t_0$}{$r$} & $1$ & $2$ & $3$ &$4$&$5$\\
		\hline
		$2$ & 5 & 15 & 34 & 76 & 150  \\
		\hline
		$3$ &13 & 69 & 211 & 596 & 1437 \\ 
		\hline
		$4$ & 45 & 315 & 1300 & 4476 &
		12996 \\ 
		\hline
		$5$ &211 & 1753 & 8659 & 35278 &
		119377 \\ 
		\hline
	\end{tabular} 
	
	\caption{Theoretical running time $T_{n,r}$ for analytical tensor reduction by direct expansion. We set $T_n[\delta D=0,V^0]=t_0$ then calculate $T_{n,r}$ through \eref{time-recur},\eref{time-dim-shift-recur} and \eref{total-time}. One can see $T_{n,r}$ suffers from sharp growth as $n,r$ increase, which is consistent with practice.  }
	\label{running time-theory}
\end{table}

    After observing the significant increase in running time, one may wonder if there is a more efficient recursive method for evaluating the integrals $I_{n;D}^{(r)}$ directly, rather than using their expansion pieces $E_{n,k}$. This question can be addressed by combining IBP and sazagy methods, as done in \cite{Chen:2022lue}. However, instead of this approach, we have derived general recursion formulas for arbitrary $n$ and $r$ directly at the integral level with the techniques in projective space.
	\section{Recursion relation for non-degenerate $Q$}
	\label{nonde-Q}
	In this section, we give the explicit expression for the recursion relation. The proof is given in Appendix \ref{appdix:lowerterm}. Then we list some examples to illustrate reduction process using the recursion relations.
	\subsection{Expression of \textit{lower term}}
	As pointed in \cite{Feng:2022rfz}, there excites a non-trivial recursion relation for one-loop tensor integrals\footnote{The formula is identical to \eref{original-form}, utilizing the fact that $L.Q^*.L=\abs{G}$ after $q_1$ is set to zero.}
	\bea
	I_{n}^{(r)}={1\over \overtr{LL}}\left[A_rI_{n}^{(r-1)}+B_rI_{n}^{(r-2)}+\lt{n}{r}\right]\ed \label{rankr-deco}
	\eea
	The two coefficients are
	\begin{align}
		A_r&={2(D+2r-n-2)\over D+r-n-1}\overtr{VL}\co \nn
		B_r&={4(r-1)(R^2-\overtr{VV})\over  D+r-n-1}\ed
	\end{align}
	Notice that due to $B_{r=1}=0$, the formula \eref{rankr-deco} works for $r\ge 1$. For $r=0$, the integral $I_n$ for general $Q$ is a master basis, so it cannot be reduced further. The \textit{lower term} $\lt{n}{r}$ corresponds to integrals with one or more propagators being canceled. In \cite{Feng:2022rfz}, we check the consistency of \eref{rankr-deco} by applying two differential operators of $R$ (\textit{i.e}., ${\cal D}_i\equiv q_{i+1}\cdot \d_R$, ${\cal T}\equiv \d_R\cdot \d_R$) on both sides. Although with the differential operators ${\cal D}_i, {\cal T}$, one can establish a reduction framework for one-loop integrals, it is quite difficult to work out $\lt{n}{r}$ with this technique for two types of complex recursion relations are involved. On the other hand, one-loop reduction in projective space only relies on one simple and symmetric recursion relation \eref{general-resur}. As shown in Appendix \ref{appdix:lowerterm}, one can figure out the expression of $\lt{n}{r}$.
	\bea
	\lt{n}{r}={\overtr{H_bL}I^{(r)}_{n;D-2,\what{b}}+ 2\overtr{H_bV}I^{(r-1)}_{n;D-2,\what{b}} \over D+r-n-1} \label{lower-d-2}
	\eea
	where we have omitted the summation over $b$ from $1$ to $n$ on the RHS. Notably, these RHS terms represent integrals in $(D - 2)$-dimensional spacetime. A reduction in $(D-2)$-dimensional spacetime is straightforward to implement:
	\begin{align}
		I_{n;D-2}^{(r)}=\sum_{j,\mathbf{b}_j}C_{n\to n;\what{\mathbf{b}}_j}^{(r)}\Big\vert_{D\to D-2}I_{n;D-2;\what{\mathbf{b}_j}}\label{d-2-reduction}
	\end{align}
	where $C_{n\to n;\what{\mathbf{b}}_j}^{(r)}$, the  reduction coefficients for $D$-dimensional integrals,  are known. In the case of non-degenerate $Q_{(b)}$, we can further reduce the lower topology term with integrals in the same dimension by reducing the right-hand side of \eref{lower-d-2}. After some algebra, we obtain
	\begin{align}
		\lt{n}{r}&=\left[\overtr{H_bL}\overtr{VL}_{(b)}-\overtr{H_bV}\overtr{LL}_{(b)}\right]I^{(r-1)}_{n;\what{b}}\nn
		&\newline+{2(r-1)\left[\overtr{H_bL}R^2+\overtr{H_bV}\overtr{VL}_{(b)}-\overtr{H_bL}\overtr{VV}_{(b)}\right]\over D+r-n-1 }I_{n;\what{b}}^{(r-2)}\nn
		&\equiv A_{r;\what{b}}I_{n;\what{b}}^{(r-1)}+B_{r;\what{b}}I_{n;\what{b}}^{(r-2)}\ed\label{lower-samed}
	\end{align}
	The details of this reduction are presented in Appendix \ref{appdix:lowerterm}, where the necessary algebraic manipulations are carried out. %The formula  is more convenient to use for we can employ reduction results of lower topologies directly.So finally we reach a more symmetric formula
	The presented formula offers superior feasibility, as a result of being able to utilize the reduction results of lower topologies directly. As such, the resultant formula achieved herein is more symmetrical in nature:
	\bea
	I_n^{(r)}={1\over \overtr{LL}}\left[A_rI_n^{(r-1)}+B_rI_n^{(r-2)}+A_{r;\what{b}}I_{n;\what{b}}^{(r-1)}+B_{r;\what{b}}I_{n;\what{b}}^{(r-2)}\right]\ed\label{recursion-nonQ}
	\eea
	The present formula contains tensor integrals with lower ranks and integrals whereby one propagator is canceled, consequently, these integrals are already known to undergo an iterative reduction process. Utilizing the recursion formula, the reduction results for any tensor integral $I_n^{(r)}$ can be obtained. The recursion diverges at $\overtr{LL}=0$, which corresponds to the singularity of the degenerate Gram matrix $\det{G}=0$\footnote{Setting $q_1=0$ reveals $\overtr{LL}=L.Q^{-1}.L=\det G/\det Q$, thereby imparting that the $\overtr{LL}=0$ pole is equal to $\det G=0$.}. The singularity is addressed simply by multiplying both sides of \eref{recursion-nonQ} with $\overtr{LL}$ and subsequently taking the limit $\overtr{LL}\to 0$,
	
	\bea
	\left[A_rI_n^{(r-1)}+B_rI_n^{(r-2)}+A_{r;\what{b}}I_{n;\what{b}}^{(r-1)}+B_{r;\what{b}}I_{n;\what{b}}^{(r-2)}\right]=0\ \text{for}\ \overtr{LL}=0\co
	\eea
	so we obtain the recursion relation for vanishing Gram determinant
	\bea
	I_{n}^{(r)}={-1\over A_{r+1}}\left[B_{r+1}I_n^{(r-1)}+A_{r+1;\what{b}}I_{n;\what{b}}^{(r)}+B_{r+1;\what{b}}I_{n;\what{b}}^{(r-1)}\right]\ed \label{deLL-recur}
	\eea
	Observe that $A_{r+1}$ contains $\overtr{VL}$ and $V_i=R\cdot q_i$. It is noteworthy that $R$ should only appear in the numerator in the final expression of reduction coefficients. Therefore, it is anticipated that the $\overtr{VL}$ in $A_{r+1}$ can be eliminated. Actually, There exist two distinct identities when $\abs{G}=0$. There are two identities for $\abs{G}=0$:
	\begin{align}
		&\overtr{H_bL}\overtr{VL}_{(b)}-\overtr{H_bV}\overtr{LL}_{(b)}=\overtr{H_bL}\overtr{VL}\co\nn
		&\overtr{H_bV}\overtr{VL}_{(b)}-\overtr{H_bL}\overtr{VV}_{(b)}+\overtr{H_bL}\overtr{VV}=\overtr{VL}\overtr{H_bV}\co
	\end{align}
	which simplify the recursion relation \eref{deLL-recur} to
	\begin{align}
		I_{n}^{(r)}&={-1\over 2(D+2r-n)}\Bigg[(D+r-n)\overtr{H_bL}I_{n;\what{b}}^{(r)}+2r\overtr{H_bV}I_{n;\what{b}}^{(r-1)}\nn&\newline +2r{R^2-\overtr{VV}\over \overtr{VL}}(2I_n^{(r-1)}
		+\overtr{H_bL}I_{n;\what{b}}^{(r-1)})\Bigg]\ed \label{deLL-simple}
	\end{align}
	It is easy to check the \textit{spurious singularity} $\overtr{VL}$ disappears in the final result for $r=0,1$:
	\begin{itemize}
		\item $r=0$
		\begin{align}
			I_n={-1\over 2}\overtr{H_bL}I_{n;\what{b}}\co
		\end{align}
		\item $r=1$
		\begin{align}
			I_n^{(1)}&={-1\over 2(D+2-n)}\Bigg[(D+1-n)\overtr{H_bL}I_{n;\what{b}}^{(1)}+2\overtr{H_bV}I_{n;\what{b}}\Bigg]\ed
		\end{align}
	\end{itemize}
	For higher rank case, one can  iteratively check $\overtr{VL}$ disappears in the denominator.
	\subsection{Running time and tensor structure of reduction coefficients}
	As stated in the introduction, utilizing recursion relations at the integrals level is a preferable approach to dealing directly with $E_{n,k}[V^i]$. Firstly, the new method exhibits notably enhanced efficiency over the previous approach. When comparing Table \underline{\ref{running time}} with Table \underline{\ref{running time-2}}, it becomes evident that the equation given in \eref{recursion-nonQ} drastically reduces the required computation time in comparison to the direct expansion method. For instance, the time required for $n=5,r=3$ in Table \underline{\ref{running time-2}} is approximately ${1/330000}$ of the time listed in Table \underline{\ref{running time}}. The reason behind this significant difference can be explained by investigating  $T_{n,r}$, which denotes the amount of time required to reduce an $r$-rank $n$-gon integral. From \eref{recursion-nonQ}, we have the following relation:
	\begin{align}
		T_{n,r}=T_{n,r-1}+T_{n,r-2}+nT_{n-1,r-1}+nT_{n-1,r-2} \label{cut-time-recur}
	\end{align} 
	which means the computational complexity of the proposed method exhibits a desirable linear relationship with respect to lower values of $n$ and $r$. For example 
	\begin{align}
		T_{5,5}&=0.635\approx T_{5,3}+T_{5,4}+5\times (T_{4,3}+T_{4,4})\nn
		&=0.00733+0.0836+5\times (0.0153+0.101)=0.672\co\nn
		T_{5,6}&=3.79\approx T_{5,4}+T_{5,5}+5\times (T_{4,4}+T_{4,5})\nn
		&=0.0836+0.635\times (0.101+0.528)=3.86\ed
	\end{align} 
	\begin{table}[h]
	\centering
	\renewcommand{\arraystretch}{1.5}
	\begin{tabular}{|>{\centering\arraybackslash}p{1.6cm}|>{\centering\arraybackslash}p{1.6cm}|>{\centering\arraybackslash}p{1.6cm}|>{\centering\arraybackslash}p{1.6cm}|>{\centering\arraybackslash}p{1.6cm}|>{\centering\arraybackslash}p{1.6cm}|>{\centering\arraybackslash}p{1.6cm}|}
		\hline
		\diagbox{$n$}{time$/s$}{$r$} & $1$ & $2$ & $3$ &$4$&$5$&$6$\\
		\hline
		$2$ & 0.000102& 0.00454& 0.0188& 0.0651& 0.181& 0.441\\
		\hline
		$3$ &0.000163& 0.00561& 0.0191&0.0921& 0.352&1.16\\ 
		\hline
		$4$ & 0.000831 &0.00263& 0.0153& 0.101&0.528&2.39\\ 
		\hline
		$5$ & 0.000955&0.00355&0.00733& 0.0836& 0.635&3.79\\
		\hline
	\end{tabular} 
	\caption{Actual running time of analytical tensor reduction by recursion formula. In comparison with Table \underline{\ref{running time}}, we restrict our focus to the non-degenerate scenario. In this context, we rely solely on the employ of the recursion formula provided by \eref{recursion-nonQ} to iteratively reduce $I_n^{(r)}$. As a result, we are able to derive results for tensor ranks of considerable magnitude within a notably brief interval of time.  }
	\label{running time-2}
\end{table}
\begin{table}[h]
	\centering
	\renewcommand{\arraystretch}{1.5}
	\begin{tabular}{|>{\centering\arraybackslash}p{1.74cm}|>{\centering\arraybackslash}p{1.6cm}|>{\centering\arraybackslash}p{1.6cm}|>{\centering\arraybackslash}p{1.6cm}|>{\centering\arraybackslash}p{1.6cm}|>{\centering\arraybackslash}p{1.6cm}|}
		\hline
		\diagbox{$n$}{time$/t_0$}{$r$} & $1$ & $2$ & $3$ &$4$&$5$\\
		\hline
		$2$ &3 & 8 & 17 & 35 & 68 \\
		\hline
		$3$ &4 & 17 & 54 & 146 & 356 \\ 
		\hline
		$4$ &5 & 26 & 115 & 425 & 1340
		 \\ 
		\hline
		$5$& 6 & 37 & 198 & 940 & 3838 
		\\ 
		\hline
	\end{tabular} 
	
	\caption{Theoretical running time of  analytical tensor reduction by recursion relation. To streamline the process, we have set $T_{n,0}$ to equal $t_0$. Subsequently, we compute $T_{n,r}$ by means of \eref{cut-time-recur}. Upon comparison with Table \underline{\ref{running time-theory}}, it becomes apparent that the rate of increase of running time is considerably slower.  }
	\label{recursion-running time-theory}
\end{table}
   %The formula \eref{cut-time-recur} is similar to \eref{time-recur}, but the key difference is that here \eref{cut-time-recur} involves the total time rather than its pieces in  \eref{time-recur}. Set $T_{n,0}=t_0$, we calculate $T_{n,r}$ in Table \underline{\ref{recursion-running time-theory}}. There is no $E_{n,k}$ expansion and dimension shifting process as before, so the reduction process can be extremely fast. To obtain numeric result, we just need to replace coefficients in the recursions with certain numbers. Similar discussion can applied to  degenerate cases we will discuss in the next section.
   The formula presented in \eref{cut-time-recur} is similar to the one in \eref{time-recur}. However, there is a significant difference between the two: \eref{cut-time-recur} involves the total time as opposed to its pieces in \eref{time-recur}. Setting $T_{n,0}=t_0$, we can calculate $T_{n,r}$ in Table \underline{\ref{recursion-running time-theory}}. The reduction process is extremely fast since there is no need for an $E_{n,k}$ expansion or a dimension shifting process. Moreover, we can quickly obtain a numeric result by replacing the coefficients in the recursions with certain numbers.

    Additionally, the recursion relation in \eref{recursion-nonQ} makes it apparent that the general structure of the reduction coefficients for arbitrary $n$ and $r$ can be studied. By observing that the singularity of reduction coefficients of a particular sector is $\overtr{LL}_{(\{j\})} \propto |G_{(\{j\})}|$, where ${j}$ is the sub-list of the label list of removed propagators, it is clear that the coefficient of the top-sector $C^{(r)}_{n\to n}$ only has the singularity of $|G|$. On the other hand, the coefficients of the next sector $C^{(r)}_{n\to n;\what{i}}$ can have the singularity of $\overtr{LL}_{(i)}\propto |G_{(i)}|$. It is interesting to note that the degree of singularity grows linearly with rank $r$ since every iteration of \eref{recursion-nonQ} lowers the rank by $1,2$ and multiplies every term by a pole factor ${1\over \overtr{LL}_{(\{j\})}}$. Ultimately, one can easily find the general structure of $C^{(r)}_{n\to n;\what{\{i\}}}$
\begin{align}
	C^{(r)}_{n\to n;\what{i_1},\what{i_2},\what{\ldots,i_m}}=\sum_{\lfloor r/2\rfloor \le k_0+k_1+\cdots+k_m\le r,k_i> 0,k_m\ge 0}{F_{i_1i_2,\ldots,i_m;k_0,k_1,\ldots,k_m} \over \overtr{LL}^{k_0}\overtr{LL}_{(i_1)}^{k_1}\overtr{LL}_{(i_1i_2)}^{k_2}\cdots \overtr{LL}_{(i_1i_2,\ldots,i_m)}^{k_m} }\ed  \label{eq:coeff-structure}
\end{align}
The structure $F_{i_1i_2,\ldots,i_m;k_0,k_1,\ldots,k_m}$ is utilized to collect contributions from $A_r,B_r,A_{r;\what{b}},B_{r;\what{b}}$ of every iteration. Hence, the function $F$ is expressed in terms of $R^2 $,$\overtr{VL}_{(\{j\})}$,$\overtr{VV}_{(\{j\})}$, $\overtr{H_bL}_{(\{j\})}$, $\overtr{H_bV}_{(\{j\})}$ where $\{j\}\subset \{i\}$. It is important to note that $F$ exhibits symmetry about $i_1,i_2,\ldots,i_m$, as demonstrated in \eref{App-B-tri-1-2}. In this paper, we present several results for rank-6 triangle to illustrate this concept.
\begin{align}
	C_{3\to 3}^{(6)}&=\frac{64 \left(D^3+15 D^2+71 D+105\right) \overtr{VL}^6}{D \left(D^2-4\right) \overtr{LL}^6}+\frac{960 \left(D^2+8 D+15\right) \overtr{VL}^4 \left(R^2-\overtr{VV}\right)}{D \left(D^2-4\right) \overtr{LL}^5}\nn
	&\newline +\frac{2880 (D+3) \overtr{VL}^2 \left(R^2-\overtr{VV}\right)^2}{D \left(D^2-4\right) \overtr{LL}^4}+\frac{960 \left(R^2-\overtr{VV}\right)^3}{D \left(D^2-4\right) \overtr{LL}^3}\co
\end{align}
\allowdisplaybreaks
\begin{align}
	C^{(6)}_{3\to 3;\what{3}}\Big\vert_{\overtr{LL}^{-6}}=\frac{32 (D+3) (D+5) (D+7) \overtr{LV}^5 \left(\overtr{LH_3} \overtr{LV}_{(3)}-\overtr{LL}_{(3)} \overtr{VH_3}\right)}{D \left(D^2-4\right)}\co
\end{align}
\begin{align}
    C^{(6)}_{3\to 3;\what{3}}\Big\vert_{\overtr{LL}^{-5}}&=-\frac{64 \left(7 D^3+45 D^2+17 D-165\right) \overtr{LL}_{\what{3}} \overtr{VH_3} \overtr{LV}^3 \left(R^2-\overtr{VV}\right)}{D \left(D^3-D^2-4 D+4\right)}\nn
    &-\frac{32 \overtr{LV}^3}{D \left(D^4-5 D^2+4\right)} \Bigg(-3 (D+5) (D+7) (D-1)^2 \overtr{LH_3} \overtr{LV}_{\what{3}} \left(R^2-\overtr{VV}\right)\nn
    &-4 D (D+1) (D+7) (D-1) \overtr{LH_3} \overtr{LV}_{\what{3}} \left(R^2-\overtr{VV}\right)\nn
    &+5 (D+1)^2 (D+3) (D-1) \overtr{LH_3} \overtr{LV}_{\what{3}} \left(\overtr{VV}-R^2\right)\nn
    &-(D+1) (D+3) (D+5) (D+7) \overtr{LH_3} \overtr{LV} \left(R^2-\overtr{VV}_{\what{3}}\right)\nn
    &-2 (D-2) (D+3) (D+5) (D+7) \overtr{LH_3} \overtr{LV}_{\what{3}} \left(R^2-\overtr{VV}\right)\nn
    &+(D-3) (D+1) (D+3) (D+5) (D+7) \overtr{VH_3} \overtr{LV}_{\what{3}} \overtr{LV}\Bigg)\nn
    &+\frac{32 (D+3) (D+5) (D+7) \overtr{LH_3} \overtr{LV}_{\what{3}}^2 \overtr{LV}^4}{D \left(D^2+D-2\right) \overtr{LL}_{\what{3}}}\co
\end{align}
\begin{align}
	C^{(6)}_{3\to 3;\what{23}}\Big\vert_{\overtr{LL}^{-5}}&=\frac{16 (D+3) (D+5) (D+7)  }{(D-1) D (D+2) }\times\nn
	&\newline\Bigg[\frac{\overtr{LH_3} \overtr{LV}_{\what{3}} \overtr{LV}^4 \left(\overtr{LV}_{\what{2,3}} \overtr{LH_2}_{\what{3}}-\overtr{LL}_{\what{2,3}} \overtr{VH_2}_{\what{3}}\right)}{ \overtr{LL}_{\what{3}}}\nn
	&\newline+\frac{ \overtr{LH_2} \overtr{LV}_{\what{2}} \overtr{LV}^4 \left(\overtr{LV}_{\what{2,3}} \overtr{LH_3}_{\what{2}}-\overtr{LL}_{\what{2,3}} \overtr{VH_3}_{\what{2}}\right)}{\overtr{LL}_{\what{2}}}\nn
	&\newline+  \overtr{LV}^4 \left(-\overtr{VH_2} \left(\overtr{LV}_{\what{2,3}} \overtr{LH_3}_{\what{2}}-\overtr{LL}_{\what{2,3}} \overtr{VH_3}_{\what{2}}\right)\right)\nn
	&\newline+ \overtr{LV}^4 \left(-\overtr{VH_3} \left(\overtr{LV}_{\what{2,3}} \overtr{LH_2}_{\what{3}}-\overtr{LL}_{\what{2,3}} \overtr{VH_2}_{\what{3}}\right)\right)\Bigg]\ed \label{Tri-Tad-Lead}
\end{align}
The equivalence of the divergence degrees between $C^{(6)}_{3\to 3}$ and $C^{(6)}_{3\to 3;\what{3}}$ for $\overtr{LL}$ is due to the fact that, for the sub-sector in which one propagator is removed, setting $k_0=6$ and $k_1=0$ in \eref{eq:coeff-structure} is permissible. However, for the sector with two propagators removed, i.e., $C^{(6)}_{3\to 3;\what{23}}$, the largest divergence degree of $\overtr{LL}$ is 5, since $k_1>0$ and $k_0+k_1+k_2\leq r=6$. The leading divergence terms correspond to $k_0=5$, $k_1=1$, and $k_0=0$. The lower degree of divergence terms are more complex, as they involve several additional contributions throughout the iteration process. Notably, the divergence of $\overtr{LL}_{(j)}$ in the sub-sector coefficients is also observed, and all of these characteristics are codified in \eref{eq:coeff-structure}. Furthermore, the symmetry of the coefficient $C^{(6)}_{3\to 3;\what{23}}\Big\vert_{\overtr{LL}^{-5}}$ with respect to the removed propagators 2 and 3 allows us to express \eref{Tri-Tad-Lead} in a simplified form.
\begin{align}
	C^{(6)}_{3\to 3;\what{23}}\Big\vert_{\overtr{LL}^{-5}}&=\frac{16 (D+3) (D+5) (D+7)  }{(D-1) D (D+2) }\times\nn
	&\newline\Bigg[\frac{\overtr{LH_3} \overtr{LV}_{\what{3}} \overtr{LV}^4 \left(\overtr{LV}_{\what{2,3}} \overtr{LH_2}_{\what{3}}-\overtr{LL}_{\what{2,3}} \overtr{VH_2}_{\what{3}}\right)}{ \overtr{LL}_{\what{3}}}\nn
	&\newline+  \overtr{LV}^4 \left(-\overtr{VH_2} \left(\overtr{LV}_{\what{2,3}} \overtr{LH_3}_{\what{2}}-\overtr{LL}_{\what{2,3}} \overtr{VH_3}_{\what{2}}\right)\right)\Bigg]+(2\leftrightarrow 3)\ed
\end{align}
The permutation symmetry appears in higher-point case, which makes the results extremely simple, as shown with more reduction results in Appendix \ref{appdix:highernr}. In Section \ref{sec:deQ}, we will discuss modified recursion relations that enable us to approach degenerate cases in a manner similar to that which we employ for non-degenerate cases. So we will not provide detailed discussion of these degenerate cases in this paper.

	%%%%%%%%%%%%%%%%%%%%%%%%%%%%%
	\subsection{Examples}
	\label{sec:nonQ-example}
	%%%%%%%%%%%%%%%%%%%%%%%%%%%%%
	In this section, we provide examples to demonstrate the use of one-loop recursion with non-degenerate $Q$. These examples serve to illustrate how the method can be applied in practice, thereby providing a deeper insight into its efficacy and versatility.
	\subsubsection{ Tadpole}
	%%%%%%%%%%%%%%%%%%%%%%%%%%%%%
	First, we consider the simplest case, \textit{i.e}, the tensor tadpole
	\bea
	I_1^{(r)}\equiv \int  {d^D \ell \over i \pi^{D/2}}  {(2R\cdot \ell)^r\over \ell^2-m_1^2}\ed
	\eea
	Due to the tadpole has one propagator, there is only one master integral $I_1$ and $\lt{1}{r}=0$. We set $q_1=0$ throughout the paper, then we find
	\begin{align}
		\overtr{LL}={1\over m_1^2},~~A_r=0,~~B_r={4(r-1)R^2\over D+r-2}\ed
	\end{align}
	The recursion relation \eref{rankr-deco} for tensor tadpole is quite simple
	\begin{align}
		I_{1}^{(r)}={4(r-1)m_1^2R^2\over D+r-2}I_1^{(r-2)}\ed
	\end{align}
	It is easy to figure out the general result
	\begin{align}
		I_{1}^{(r)}=\left\{\begin{matrix}
			&0\ ,&r= \text{odd}\\ &{2^r(r-1)!!(m_1^rR^r)\over (D+r-2)!!}I_1,~~&r= \text{even}
		\end{matrix}\right.
	\end{align}
	\subsubsection{ Bubble}
	To avoid unnecessary complicate calculations and to show the advantages of reduction in projective space, we consider tensor bubbles as a  non-trivial example
	\bea
	I_2^{(r)}\equiv \int {d^D \ell \over i \pi^{D/2}} {(2R\cdot \ell)^r\over (\ell-m_1^2)[(\ell-q_2)^2-m_2^2]}
	\eea
	where we have set $q_1=0$. The recursion relation reads
	\bea
	I_2^{(r)}={1\over \overtr{LL}}\left[A_rI_2^{(r-1)}+B_rI_2^{(r-2)}+A_{r;\what{b}}I_{2;\what{b}}^{(r-1)}+B_{r;\what{b}}I_{2;\what{b}}^{(r-2)}\right]\label{bubble-recur}\ed
	\eea
	The recursion gives the reduction coefficients written in a compact form
	\bea
	C^{(1)}_{2\to 2;\widehat{i}}&=&-\frac{\overtr{L L} \overtr{H_i V}-\overtr{H_i L} \overtr{V L}}{\overtr{L L}}\co~~
	C^{(1)}_{2\to 2}
	=\frac{2 \left(VQ^{-1}L\right)}{\left(LQ^{-1}L\right)}\co\nn
    C^{(2)}_{2\to 2;{\widehat{i}}}&=&\frac{2 R^2 \overtr{H_i L}}{(D-1) \overtr{L L}}+\frac{2 D \overtr{H_i L} \overtr{V L}^2}{(D-1) \overtr{L L}^2}-\frac{2 \overtr{H_i V} \left(\overtr{V L}+\overtr{V L}_{(i)}\right)}{\overtr{L L}_{(i)}}\nn
    &&-\frac{2 \overtr{H_i L} \big(-D \overtr{V L}^2+\overtr{V L}^2+\overtr{L L}_{(i)} \overtr{V V}\big)}{(D-1) \overtr{L L}_{(i)} \overtr{L L}}\co\nn
    C^{(2)}_{2\to 2}
    &=&\frac{4 \left(R^2-\overtr{V V}\right)}{(D-1) \overtr{L L}}+\frac{4 D \overtr{V L}^2}{(D-1) \overtr{L L}^2}\ed~~~~\label{bub-re}
    \eea
	To simplify notation, we use $R^2=s_{00}$, $R\cdot q_2=s_{01}$, and $q_2^2=s_{11}$ for the bubble. By performing direct calculations, we obtain the matrix $Q$
	\begin{align}
		Q&=\left(
		\begin{array}{cc}
			m_1^2 & {\frac{1}{2} \left(m_1^2+m_2^2-s_{11}\right)} \\
			{\frac{1}{2} \left(m_1^2+m_2^2-s_{11}\right)} & m_2^2 \\
		\end{array}
		\right)\co\nn
		Q^{-1}&={-1\over 4|Q|}\left (
			\begin{array}{cc}
				-4 m_2^2 &
				2 \left(m_1^2+m_2^2-s_{11}\right) \\
				2 \left(m_1^2+m_2^2-s_{11}\right) &
				-4 m_1^2 \\
			\end{array}
			\right)
%\over -2 m_1^2\left(m_2^2+s_{11}\right)+\left(m_2^2-s_{11}\right){}^2+m_1^4}
\ed
	\end{align}
	And the sub-matrices of $Q$ are
	\begin{align}
		Q_{(1)}=\left(
		\begin{array}{cc}
			\xcancel{Q_{11}} & \xcancel{Q_{12}} \\
			\xcancel{Q_{21}} & {Q_{22}} \\
		\end{array}
		\right)=m_2^2,~~
		Q_{(2)}=\left(
		\begin{array}{cc}
			Q_{11} & \xcancel{Q_{12}} \\
			\xcancel{Q_{21}} & \xcancel{Q_{22}} \\
		\end{array}
		\right)=m_1^2\ed
	\end{align}
	With the expressions for $Q$ and its sub-matrices, it becomes effortless to compute these concise cells using the recursive relationship \eref{bubble-recur}, as illustrated below:
	\begin{align}
		\overtr{H_1L}&\equiv H_1Q^{-1}L=\frac{2 \left(m_1^2-m_2^2-s_{11}\right)}{-2 m_1^2
			\left(m_2^2+s_{11}\right)+\left(m_2^2-s_{11}\right){}^2+m_1^4}\co\nn
		\overtr{H_2L}&\equiv H_2Q^{-1}L=-\frac{2 \left(m_1^2-m_2^2+s_{11}\right)}{-2 m_1^2
			\left(m_2^2+s_{11}\right)+\left(m_2^2-s_{11}\right){}^2+m_1^4}\ed
	\end{align}
	By substituting the aforementioned expressions into the coefficients involved in \eref{bubble-recur}, one can obtain the final result as follows:
	\begin{align}
		I_2^{(r)}&=\frac{(D+2r-4)\left(m_1^2-m_2^2+s_{11}\right) s_{01}}{(D+r-3) s_{11}}I_2^{(r-1)}+\frac{s_{01}}{s_{11}}I_{2;\what{1}}^{(r-1)}-\frac{s_{01}}{s_{11}}I_{2;\what{2}}^{(r-1)}\nn
		&\newline+{r-1\over (D+r-3)s_{11}}\Bigg[ -\left[ \left(4 m_1^2 s_{01}^2+\left(-2 m_1^2
		\left(m_2^2+s_{11}\right)+\left(m_2^2-s_{11}\right){}^2+m_1^4\right)
		s_{00}\right)\right]I_2^{(r-2)}\nn
		&\newline + \left(\left(-m_1^2+m_2^2+s_{11}\right) s_{00}-2
		s_{01}^2\right)I_{2;\what{1}}^{(r-2)}+ \left(m_1^2-m_2^2+s_{11}\right) s_{00}I_{2;\what{2}}^{(r-2)}\Bigg]\ed \label{bubble=recur-detail}
	\end{align}
	Using the reduction results for tensor tadpoles, it becomes possible to iteratively apply the recursion relation to reduce a tensor bubble up to any desired rank. This can be illustrated by the following example:
	\allowdisplaybreaks
	\begin{itemize}
		\item $r=1$
		\begin{align}
			I_2^{(1)}&=\frac{\left(m_1^2-m_2^2+s_{11}\right) s_{01}}{ s_{11}}I_2+\frac{s_{01}}{s_{11}}I_{2;\what{1}}-\frac{s_{01}}{s_{11}}I_{2;\what{2}}\ed
		\end{align}
		\item $r=2$
		\begin{align}
			I_2^{(2)}&=\frac{D\left(m_1^2-m_2^2+s_{11}\right) s_{01}}{(D-1) s_{11}}I_2^{(1)}+\frac{s_{01}}{s_{11}}I_{2;\what{1}}^{(1)}-\frac{s_{01}}{s_{11}}I_{2;\what{2}}^{(1)}\nn
			&\newline+{1\over (D-1)s_{11}}\Bigg[ -\left[ \left(4 m_1^2 s_{01}^2+\left(-2 m_1^2
			\left(m_2^2+s_{11}\right)+\left(m_2^2-s_{11}\right){}^2+m_1^4\right)
			s_{00}\right)\right]I_2\nn
			&\newline + \left(\left(-m_1^2+m_2^2+s_{11}\right) s_{00}-2
			s_{01}^2\right)I_{2;\what{1}}+ \left(m_1^2-m_2^2+s_{11}\right) s_{00}I_{2;\what{2}}\Bigg]\ed
		\end{align}
		Then use the results of rank-$1$ bubble reduction, we finally find
		\begin{align}
			I_2^{(2)}&=\Bigg[\frac{D \left(m_1^2-m_2^2\right){}^2 s_{01}^2}{(D-1) s_{11}^2}-\frac{-2 m_1^2 \left((D-2) s_{01}^2+m_2^2 s_{00}\right)+2 D m_2^2 s_{01}^2+m_1^4 s_{00}+m_2^4 s_{00}}{(D-1) s_{11}}\nn
			&\newline -\frac{s_{11} s_{00}}{D-1}+\frac{D s_{01}^2+2 m_1^2 s_{00}+2 m_2^2 s_{00}}{D-1}\Bigg]I_2+\frac{\left(m_1^2-m_2^2+s_{11}\right)  \left(s_{11} s_{00}-D s_{01}^2\right)}{(D-1) s_{11}^2}I_{2;\what{2}}\nn&\newline+\frac{\left(s_{01}^2 \left(D m_1^2-D m_2^2+(3 D-4) s_{11}\right)+s_{11} \left(-m_1^2+m_2^2+s_{11}\right) s_{00}\right)}{(D-1) s_{11}^2}I_{2;\what{1}}\ed
		\end{align}
	\end{itemize}
	These results agree with that provided by PV reduction. But comparing \eref{bubble=recur-detail} with its compact form  \eref{bub-re}, one can see  analytical structure and  permutation symmetries of propagators manifest in projective space language. Here, consider the permutation exchanging the two propagators of bubble: $\sigma(1)=2,\sigma(2)=1$. From \eref{bub-re}, we find
	\begin{align}
		\sigma [C^{(r=1,2)}_{2;\what{2}}]=C^{(r=1,2)}_{2;\what{1}}=C^{(r=1,2)}_{2;\what{\sigma(2)}}\ed
	\end{align}
	Same things happen for higher $n,r$. Moreover, reduction recursion relies on $n$ trivially for $n$-gon tensor integrals. for example, one can easily obtain the reduction coefficients of a rank-$1$ triangle
	\bea
	\tensorc{1}{3}{i}&=&\frac{\overtr{H_i L} \overtr{V L}}{\overtr{L L}}-\overtr{H_i V}\co\nn
	\tensorselfc{1}{3}&=&\frac{2 \overtr{V L}}{\overtr{L L}}\ed
	\eea
	It has the same form as rank-$1$ bubble.

	 One can notice that there is a singularity $s_{11}=q_2^2=0$ in the recursion relation \eref{bubble=recur-detail}, which comes from $\overtr{LL}=0$ in \eref{bubble-recur}. We address the singularity by \eref{deLL-recur}
	\bea
	I_{2}^{(r)}={-1\over A_{r+1}}\left[B_{r+1}I_2^{(r-1)}+A_{r+1;\what{b}}I_{2;\what{b}}^{(r)}+B_{r+1;
\what{b}}I_{2;\what{b}}^{(r-1)}\right]\ed
	\eea
	Using
	\begin{align}
		Q&=\left(
		\begin{array}{cc}
			m_1^2 & \frac{1}{2} \left(m_1^2+m_2^2\right) \\
			\frac{1}{2} \left(m_1^2+m_2^2\right) & m_2^2 \\
		\end{array}
		\right)\co\nn
		Q^{-1}&={1\over (m_1-m_2)^2}\left (
		\begin{array}{cc}
			-4 m_2^2 & 2 \left(m_1^2+m_2^2\right) \\
			2 \left(m_1^2+m_2^2\right) & -4 m_1^2 \\
		\end{array}
		\right)\co
	\end{align}
	we finally find
	\begin{align}
		A_{r+1}&={2(D+2r-2)\over D+r-2},~\overtr{VL}=-\frac{4 s_{01} (D+2 r-2)}{\left(m_1^2-m_2^2\right) (D+r-2)}\co\nn
		B_{r+1}&={4r(R^2-\overtr{VV})\over  D+r-2}=\frac{4 r \left(\frac{4 m_1^2 s_{01}^2}{\left(m_1^2-m_2^2\right){}^2}+s_{00}\right)}{D+r-2}\co\nn
		A_{r+1;\what{b}}&=\overtr{H_bL}\overtr{VL}_{(b)}-\overtr{H_bV}\overtr{LL}_{(b)}=\left\{-\frac{4 s_{01}}{\left(m_1^2-m_2^2\right){}^2},\frac{4 s_{01}}{\left(m_1^2-m_2^2\right){}^2}\right\}\co\nn
		B_{r+1;\what{b}}&={2r\left[\overtr{H_bL}R^2+\overtr{H_bV}\overtr{VL}_{(b)}-\overtr{H_bL}\overtr{VV}_{(b)}\right]\over D+r-2}\nn
		&={2r\over D+r-2}\left\{\frac{2 \left(m_1^2-m_2^2\right) s_{00}^2+4 s_{01}^2}{\left(m_1^2-m_2^2\right){}^2},-\frac{2 s_{00}}{m_1^2-m_2^2}\right\}\ed
	\end{align}
	So we have
	\begin{align}
		I_{2}^{(r)}
		&={1\over s_{01}(D+2r-2)}\Bigg[r \left(\frac{4 m_1^2 s_{01}^2}{\left(m_1^2-m_2^2\right)}+s_{00}\left(m_1^2-m_2^2\right)\right)I_2^{(r-1)}-\frac{(D+r-2) s_{01}}{\left(m_1^2-m_2^2\right)}I_{2;\what{1}}^{(r)}\nn
		&\newline+\frac{(D+r-2) s_{01}}{\left(m_1^2-m_2^2\right)}I_{2;\what{2}}^{(r)} +\frac{r \left(m_1^2-m_2^2\right) s_{00}^2+2r s_{01}^2}{\left(m_1^2-m_2^2\right)}I_{2;\what{1}}^{(r-1)}-r s_{00}I_{2;\what{2}}^{(r-1)}\Bigg]\ed
	\end{align}
	We can utilize this recursion to iteratively reduce a tensor bubble, as illustrated below:
	\begin{itemize}
		\item $r=0$
		\begin{align}
			I_2&=-\frac{1}{\left(m_1^2-m_2^2\right)}I_{2;\what{1}}+\frac{1}{\left(m_1^2-m_2^2\right)}I_{2;\what{2}}\ed
		\end{align}
		\item $r=1$
		\begin{align}
			I_2^{(1)}&={1\over s_{01}D}\Bigg[ \left(\frac{4 m_1^2 s_{01}^2}{\left(m_1^2-m_2^2\right)}+s_{00}\left(m_1^2-m_2^2\right)\right)I_2-\frac{(D-1) s_{01}}{\left(m_1^2-m_2^2\right)}I_{2;\what{1}}^{(1)}\nn
			&\newline+\frac{(D-1) s_{01}}{\left(m_1^2-m_2^2\right)}I_{2;\what{2}}^{(1)} +\frac{ \left(m_1^2-m_2^2\right) s_{00}+2 s_{01}^2}{\left(m_1^2-m_2^2\right)}I_{2;\what{1}}- s_{00}I_{2;\what{2}}\Bigg]\ed
		\end{align}
		Then using the result we have got
		\begin{align}
			I_2&=-\frac{1}{\left(m_1^2-m_2^2\right)}I_{2;\what{1}}+\frac{1}{\left(m_1^2-m_2^2\right)}I_{2;\what{2}},I_{2;\what{1}}^{(1)}=2s_{01}I_{2;\what{1}},I_{2;\what{2}}^{(1)}=0\co
		\end{align}
		we finally find
		\begin{align}
			I_2^{(1)}&={1\over s_{01}D}\Bigg[ \left(\frac{4 m_1^2 s_{01}^2}{\left(m_1^2-m_2^2\right)}+s_{00}\left(m_1^2-m_2^2\right)\right)\left(-\frac{1}{\left(m_1^2-m_2^2\right)}I_{2;\what{1}}+\frac{1}{\left(m_1^2-m_2^2\right)}I_{2;\what{2}}\right)\nn
			&\newline -\frac{(D-1) s_{01}}{\left(m_1^2-m_2^2\right)}2s_{01}I_{2;\what{1}}+\frac{ \left(m_1^2-m_2^2\right) s_{00}+2 s_{01}^2}{\left(m_1^2-m_2^2\right)}I_{2;\what{1}}- s_{00}I_{2;\what{2}}\Bigg]\nn
			&=-\frac{2 \left(D m_1^2-(D-2) m_2^2\right) s_{01}}{D \left(m_1^2-m_2^2\right)^2}I_{2;\what{1}}+\frac{4 m_1^2 s_{01}}{D \left(m_1^2-m_2^2\right)^2}I_{2;\what{2}}\ed
		\end{align}
	\end{itemize}
	Moreover, there exists another singularity, specifically when $m_1=m_2$, within the recursion relation. In such an instance, the associated matrix $Q$ becomes degenerate, and it will be analyzed in the next section.
	
	\subsubsection{ Triangle}
	Next we consider a nontrivial case, the triangle with 
	\bea
	I_3^{(r)}\equiv \int {d^D \ell \over i \pi^{D/2}} {(2R\cdot \ell)^r\over (\ell-m_1^2)[(\ell-q_2)^2-m_2^2][(\ell-q_3)^2-m_3^2]}\ed
	\eea
    We denote $R^2=s_{00},R\cdot q_i=s_{0(i-1)},q_i\cdot q_j=s_{(i-1)(j-1)}$ and write the reduction results as 
    \begin{align}
    	I_3^{(r)}=\tensorselfc{r}{3}I_3+\sum_{1\le i \le 3}\tensorc{r}{3}{i}I_{3;\what{i}}+\sum_{1\le i<j\le 3}\tensorc{r}{3}{ij}I_{3;\what{ij}}\ed
    \end{align} 
    By simple iteration, we can find the reduction result for rank $r=1,2$
    \bea
    \tensorc{1}{3}{i}=\frac{\overtr{H_i L} \overtr{V L}}{\overtr{L L}}-\overtr{H_i V},~~\tensorselfc{1}{3}=\frac{2 \overtr{V L}}{\overtr{L L}}\ed
    ~~~~\label{App-B-tri-1-1}
    \eea
    \bea
    \tensorc{2}{3}{ij}&=&\frac{\overtr{H_i L} \overtr{H_j L}_{(i)} \overtr{V L}^2}{\overtr{L L}_{(i)} \overtr{L L}}-\frac{\overtr{H_j L}_{(i)} \overtr{H_i V} \left(\overtr{V L}_{(i)}+\overtr{V L}\right)}{\overtr{L L}_{(i)}}\nn
    &&+\overtr{H_i V} \overtr{H_j V}_{(i)}+(i\leftrightarrow j)\co\nn
    \tensorc{2}{3}{i}&=&\frac{2 \overtr{H_i L} \left((D-2) \overtr{V L}^2+\overtr{L L}_{(i)} \left(R^2-\overtr{V V}\right)\right)}{(D-2) \overtr{L L}_{(i)} \overtr{L L}}\nn
    &&+\frac{2 (D-1) \overtr{H_i L} \overtr{V L}^2}{(D-2) \overtr{L L}^2}\-\frac{2 \overtr{H_i V} \left(\overtr{V L}_{(i)}+\overtr{V L}\right)}{\overtr{L L}_{(i)}}\co\nn
    \tensorselfc{2}{3}&=&\frac{4 \left((D-1) \overtr{V L}^2+\overtr{L L} \left(R^2-\overtr{V V}\right)\right)}{(D-2) \overtr{L L}^2}\ed~~~~\label{App-B-tri-1-2}
    \eea
    By direct calculation, one finds the  $Q$ matrix are given by
    \begin{align}
    	Q&=\left(
    	\begin{array}{ccc}
    		m_1^2 & \frac{1}{2} \left(m_1^2+m_2^2-s_{11}\right) & \frac{1}{2}
    		\left(m_1^2+m_3^2-s_{22}\right) \\
    		\frac{1}{2} \left(m_1^2+m_2^2-s_{11}\right) & m_2^2 & \frac{1}{2}
    		\left(m_2^2+m_3^2-s_{11}-s_{22}\right)+s_{12} \\
    		\frac{1}{2} \left(m_1^2+m_3^2-s_{22}\right) & \frac{1}{2}
    		\left(m_2^2+m_3^2-s_{11}-s_{22}\right)+s_{12} & m_3^2 \\
    	\end{array}
    	\right)\ed
    \end{align}  
    One can find that $\overtr{LL}\equiv L.Q^{-1}.L={\det G\over \det Q}$ and obtain the expression of reduction coefficients in terms of Lorentz products and masses, for example
    \begin{align}
    	\tensorc{1}{3}{3}&=\frac{s_{12} s_{01}}{s_{11} s_{22}-s_{12}^2}+\frac{s_{11} s_{02}}{s_{12}^2-s_{11} s_{22}}\co\nn
    	\tensorselfc{1}{3}&=\frac{\left(s_{22} \left(m_1^2-m_2^2+s_{11}\right)-s_{12} \left(m_1^2-m_3^2+s_{22}\right)\right) s_{01}}{s_{11} s_{22}-s_{12}^2}\nn
    	&\newline +\frac{\left(s_{11} \left(m_1^2-m_3^2+s_{22}\right)-s_{12} \left(m_1^2-m_2^2+s_{11}\right)\right) s_{02}}{s_{11} s_{22}-s_{12}^2}\ed
    \end{align} 
    It is obvious that there is a singularity at $\det G=0$ in these expressions.  To make our expression simple and the singularity  appear obviously, we can take $\{m_1^2,m_2^2,m_3^2\}=\{1/2,1/3,1/5\},s_{11}=5/7,s_{12}=7/13,s_{22}=\frac{7 t}{5}+\frac{343}{845}$. Then $\det G=t$ and the result of rank $r=2$ are
\begin{align}
	\tensorc{2}{3}{23}&=\frac{-2401 s_{01}^2+6370 s_{02} s_{01}-4225 s_{02}^2}{3185 t}+{\cal O}\left(t^0\right)\co\nn
	\tensorc{2}{3}{3}&=\frac{24787 s_{01}^2-5200 s_{02} s_{01}-38025 s_{02}^2+530 s_{00}}{35490 t}-\frac{53 \left(49 s_{01}-65 s_{02}\right){}^2}{6997445 t^2}+{\cal O}\left(t^0\right)\co\nn
	\tensorselfc{2}{3}&=\frac{-476472423 s_{01}^2+1084364190 s_{02} s_{01}-596126375 s_{02}^2-1573040 s_{00}}{2519080200 t}\nn
	&\newline +\frac{5618 \left(49 s_{01}-65 s_{02}\right){}^2}{17738523075 t^2}+{\cal O}\left(t^0\right)\ed
\end{align}
 Then we use \eref{deLL-simple} to address the singularity at $t=0$
\begin{align}
		I_{3}^{(r)}&={-1\over 2(D+2r-3)}\Bigg[(D+r-3)\overtr{H_bL}I_{3;\what{b}}^{(r)}+2r\overtr{H_bV}I_{3;\what{b}}^{(r-1)}\nn&\newline +2r{(R^2-\overtr{VV})\over \overtr{VL}}(2I_3^{(r-1)}
	+\overtr{H_bL}I_{3;\what{b}}^{(r-1)})\Bigg]
\end{align}
where
\begin{align}
	Q=\left(
	\begin{array}{ccc}
		\frac{1}{2} & \frac{5}{84} & \frac{497}{3380} \\
		\frac{5}{84} & \frac{1}{3} & \frac{4348}{17745} \\
		\frac{497}{3380} & \frac{4348}{17745} & \frac{1}{5} \\
	\end{array}
	\right),\det Q=-\frac{2809}{8996715}\not =0\co
\end{align}
and 
\begin{align}
	\overtr{H_bL}&=\Bigg\{-\frac{18720}{3649},-\frac{57330}{3649},\frac{76050}{3649}\Bigg\},~~
	\overtr{H_bV}=\Bigg\{\frac{55690092575 s_{02}-48071519811 s_{01}}{1864128140},\nn
	&\newline -\frac{27 \left(1909403979 s_{01}-2306947175 s_{02}\right)}{1065216080},\frac{845 \left(103198347 s_{01}-119140775 s_{02}\right)}{1491302512}\Bigg\}\ed
\end{align}
You can find for rank $r=0$, the scalar triangle degenerates to three bubbles
\begin{align}
	I_3=\frac{2535}{106} I_{3;\what{3}}-\frac{1911}{106} I_{3;\what{2}}-\frac{312}{53} I_{3;\what{1}}\ed
\end{align}
For rank $r=1,2,3$, we list the necessary results while other coefficients can be obtained by permutations of propagators. All results are checked with Fire6.
\begin{align}
	C_{3\to 3;\what{23}}^{(1)}&=-\frac{169}{371} \left(49 s_{01}-65 s_{02}\right),~~ C_{3\to 3;\what{3}}^{(1)}=\frac{169 \left(972595 s_{02}-568463 s_{01}\right)}{943824}\co\nn
	\tensorc{2}{3}{23}&=-\frac{13 \left(-33880938378 s_{01}^2+111907128060 s_{02} s_{01}-90727388050 s_{02}^2+192697400 s_{00}\right)}{4046645400}\co\nn
	\tensorc{2}{3}{3}&=\frac{169 \left(352548689829 s_{01}^2-1159171789230 s_{02} s_{01}+5755 \left(164368555 s_{02}^2+786520 s_{00}\right)\right)}{105047611200}\co\nn
	\tensorc{3}{3}{23}&=-\frac{56902570778731 s_{01}^3}{87539676000}+\frac{2720957887418641 s_{02} s_{01}^2}{857888824800}+\frac{90261977 s_{00} s_{02}}{7635180}\nn  &\newline +\left(-\frac{8594887965495949 s_{02}^2}{1681462096608}-\frac{10624609 s_{00}}{779100}\right) s_{01}
	+\frac{226385329727449243 s_{02}^3}{82391642733792}\co\nn
	\tensorc{2}{3}{3}&=-\frac{15431442320102783 s_{01}^3}{4544917056000}+\frac{104751998213085899 s_{02} s_{01}^2}{6362883878400}+\frac{37837641361 s_{00} s_{02}}{396406080}\nn &\newline +\left(-\frac{47377305338576681 s_{02}^2}{1781607485952}-\frac{12745079399 s_{00}}{283147200}\right) s_{01}+\frac{1243863687023160671 s_{02}^3}{87298766811648}\ed
\end{align}
\subsubsection{ Hexagon} 
We consider $D=4-2\eps$ and all external momenta are in $4-$dimensional spacetime. One can prove that the matrix $Q$ for a hexagon isn't degenerate for general moment. But here we have $LQ^{-1}L=0$. So using \eref{deLL-simple}, we find
	\begin{itemize}
		\item $r=0$
		
		\begin{align}
			I_6=-{1\over 2}\overtr{H_bL}I_{6;\what{b}}\co
		\end{align}
		
		\item $r=1$
		\begin{align}
			I_{6}^{(1)}&={-1\over 2(D-4)}\Bigg[(D-5)\overtr{H_bL}I_{6;\what{b}}^{(1)}+2\overtr{H_bV}I_{6;\what{b}}\Bigg]\ed
		\end{align}
		
	\end{itemize}
	For  $n>6$, the matrix $Q$ is degenerate, we discuss it in the next section.

	\section{Recursion of degenerate $Q$}
	\label{sec:deQ}
	
	In the preceding section, we presented a reduction for non-degenerate $Q$ based on the formula presented in equation \eref{rankr-deco}. However, there are certain cases where the recursion in equation \eref{rankr-deco} breaks down. Specifically, there are two cases: (A) when the matrix $Q$ degenerates, rendering expressions such as $\overtr{WZ}\equiv W.Q^{-1}.Z$ ill-defined, and (B) when $Q$ is non-degenerate but $\overtr{LL}=0$, which has been addressed in equation \eref{deLL-recur}.
	
	In this section, our focus primarily lies on adjusting the formula in equation \eref{rankr-deco} to accommodate for degenerate $Q$. Then we give some examples to demonstrate how the aforementioned modification can be applied in practical scenarios, thereby providing a deeper understanding of the efficacy and versatility of the modified formula.
	\subsection{Modified recursion relations}
	In the context of practical calculations of scattering amplitudes, it is apparent that there may be instances where the matrix $Q$ becomes degenerate. This typically occurs when working in finite dimensions while considering special kinetic configurations. Under such circumstances, our recursion \eref{rankr-deco} may break down. For instance, in the dimension regularization scheme, where $D=d-2\eps$, and the external momenta exist in d-dimensional spacetime, the matrix $Q$ must be degenerate if $n$ exceeds $d+2$. Furthermore, certain momenta or mass configurations can also lead to a degenerate matrix $Q$. For example, in a scattering process involving photons or gluons, some external momenta or inner propagators may be massless, leading to a degenerate $Q$. We provide a derivation of \eref{rankr-deco} in Appendix \ref{appdix:lowerterm}, which can also be adapted to cover degenerate $Q$. The only modification required is to replace $Q^{-1}$ with an arbitrary symmetric matrix $\widetilde{Q}$ in the intermediate steps. Ultimately, this leads to a general recursion relation with every term containing a $\widetilde{Q}$.
	\bea
	\adtr{LL}I_{n}^{(r)}=\left[\wt{A}_rI_{n}^{(r-1)}+\wt{B}_r^{(RR)}I_{n}^{(r-2)}[Q\wt{Q}]+\wt{B}_r^{(VV)}I_{n}^{(r-2)}+\wt{A}_{r;\what{b}}I^{(r-1)}_{n;\what{b}}+\wt{B}_{r;\what{b}}I^{(r-2)}_{n;\what{b}}\right]\ed \label{general-rankr-deco}
	\eea
	where  we have defined $\adtr{WZ}\equiv \sum_{I,J=1}^nW^I\wt{Q}_{IJ}Z^J$.
	 The coefficients in \eref{general-rankr-deco} are
	\bea
	\wt{A}_r={2(D+2r-n-2)\over D+r-n-1}\adtr{VL},~~\wt{B}_r^{(RR)}={4(r-1)R^2\over  D+r-n-1},~~\wt{B}_r^{(VV)}={-4(r-1)\adtr{VV}\over  D+r-n-1}
	\eea
	\begin{align}
		\wt{A}_{r;\widehat{b}}&=\left[\adtr{H_bL}\overtr{VL}_{(b)}-\adtr{H_bV}\overtr{LL}_{(b)}\right]\co\nn
		\wt{B}_{r;\widehat{b}}&={2(r-1)\left[\adtr{H_bL}R^2+\adtr{H_bV}\overtr{VL}_{(b)}-\adtr{H_bL}\overtr{VV}_{(b)}\right]\over D+r-n-1 }
	\end{align}
	where we split the coefficient $\widetilde{B}_r$ into two parts $\wt{B}_r^{(RR)}$ and $\wt{B}_r^{(VV)}$ and the first one contains no $\widetilde{Q}$. To keep the homogeneous condition for $\widetilde{Q}$, we need to add a $\widetilde{Q}$ into $I_n^{(r)}[Q\widetilde{Q}]$ in \eref{general-rankr-deco}
	\bea I^{(r)}_n[Q\wt{Q}]&= &{\Gamma(n-D/2-r)\over (-)^{n+r}}\sum_{i=0}^{r}\mathscr{C}^{D/2+r-n}_{r,i}
	(R^2)^{r-i\over 2} E_{n,n-D-r+i}[(Q\wt{Q}V)\otimes V^{i-1}]
	\eea
	where $(Q\wt{Q}V)_{i}= Q_{ij}\wt{Q}^{jk}V_k$ is a vector. Notice that in the derivation of \eref{general-rankr-deco} we  assume $Q_{(b)}$, \textit{i.e.}, the sub-matrix of $Q$, is non-degenerate. If not, we need to start with \eref{lower-d-2} and introduce another $(n-1)\times (n-1)$ general matrices $\widetilde{Q_{(b)}}$ to reduce $\lt{n}{r}$.
	
	It is obvious that \eref{general-rankr-deco} can return to the non-degenerate case by choose $\widetilde{Q}=Q^{-1}$. For degenerate $Q$, we can choose a $\widetilde{Q}$ satisfying $Q\wt{Q}=0$, then the second term $\wt{B}_r^{(RR)}I_{n}^{(r-2)}[Q\wt{Q}]$ in \eref{general-rankr-deco} vanishes, and \eref{general-rankr-deco} becomes
	\bea	\adtr{LL}I_{n}^{(r)}=\left[\wt{A}_rI_{n}^{(r-1)}+\wt{B}_r^{(VV)}I_{n}^{(r-2)}+\wt{A}_{r;\what{b}}I^{(r-1)}_{n;\what{b}}+\wt{B}_{r;\what{b}}I^{(r-2)}_{n;\what{b}}\right]\ed \label{rankr-deco-2}
	\eea
	Similar to the non-degenerate $Q$, depending on whether $\adtr{LL}$ is zero or not, we have two cases to deal with:
	\begin{itemize}
		\item $\adtr{LL}\not =0$
		
		We fist consider the simple case $\adtr{LL}\not =0$, then we get a recursion relation similar to \eref{recursion-nonQ} but without the coefficient $\widetilde{B}_r^{(RR)}$
		\begin{align}
			I_{n}^{(r)}&={1\over \adtr{LL}}\left[\wt{A}_rI_{n}^{(r-1)}+\wt{B}_r^{(VV)}I_{n}^{(r-2)}+\wt{A}_{r;\what{b}}I^{(r-1)}_{n;\what{b}}+\wt{B}_{r;\what{b}}I^{(r-2)}_{n;\what{b}}\right]\ed \label{deLLnot}
		\end{align}
		This formula is valid for $r\ge 1$. For $r=0$, from the discussion in \cite{Feng:2022rwj}, $I_n$ doesn't belong to the basis anymore.
	    One can first write $I_n=(-1)^n{\Gamma(n-D/2)}E_{n,n-D}$, then applying the recursion relation for $E_{n,n-D}$
		\begin{align}
			E_{n,n-D}={-\adtr{H_bL}\over (n-D+1)\adtr{LL}}E_{n-1,n-D+1}^{(b)}\ed
		\end{align}
	    The $E^{(b)}_{n-1,n-D+1}$ can be reduced further depending on whether $Q_{(b)}$ is degenerate or not.
		
		\item $\adtr{LL}=0$
		
		Now \eref{rankr-deco-2} becomes
		\bea
		0=\left[\wt{A}_rI_{n}^{(r-1)}+\wt{B}_r^{(VV)}I_{n}^{(r-2)}+\wt{A}_{r;\what{b}}I^{(r-1)}_{n;\what{b}}+\wt{B}_{r;\what{b}}I^{(r-2)}_{n;\what{b}}\right] \label{rankr-deco-LLzero}\ed
		\eea
		
		Then there are two subcases:
		\begin{itemize}
			\item $\wt{Q}L\not =0$:
			The coefficient $\widetilde{A}_r$ is nonzero, so
			\begin{align}
			I_n^{(r)}={-1\over \wt{A}_{r+1}}\left[\wt{B}_{r+1}^{(VV)}I_{n}^{(r-1)}+\wt{A}_{r+1;\what{b}}I^{(r)}_{n;\what{b}}+\wt{B}_{r+1;\what{b}}I^{(r-1)}_{n;\what{b}}\right]\ed
			\end{align}
			\item $\wt{Q}L =0$: The coefficient $\widetilde{A}_r$ vanishes, so finally we get
			\begin{align}
				I_n^{(r)}
				&={\adtr{H_bV}\over \wt{B}_{r+2}^{(VV)}}\left[\overtr{LL}_{(b)}I^{(r+1)}_{n;\what{b}}-{2(r+1)\overtr{VL}_{(b)}\over D+r-n+1}I^{(r)}_{n;\what{b}}\right]\ed\label{deLLdeVL}
			\end{align}
			
		\end{itemize}
		Both two recursion relations work for $r\ge 0$.
	\end{itemize}
	At last, we consider a very special case where the sub-matrices $Q_{(b)}$ degenerate too, which makes some coefficients in \eref{general-rankr-deco} diverge. one need to use the $(D-2)$-dimensional lower terms
	\begin{align}
		\lt{n}{r}={\overtr{H_bL}I^{(r)}_{n;D-2,\what{b}}+ 2\overtr{H_bV}I^{(r-1)}_{n;D-2,\what{b}} \over D+r-n-1}\co
	\end{align}
	then reduce these terms by \eref{d-2-reduction}. Another method is to expand reduction coefficients and some degenerate basis according to $\abs{Q_{(b)}}$, then taking the limit $\abs{Q_{(b)}}\to 0$ (see \cite{Feng:2022rfz}).

	\subsection{Examples}
	Here we give some examples to illustrate how to apply the recursion  relations in this section  to reduction for degenerate cases. To compare with the non-degenerate case that we have discussed in Section \ref{sec:nonQ-example}, here we mainly deal with bubbles and triangles.
	\subsubsection{Bubble}
	\begin{itemize}
		\item
		\textbf{$L\widetilde{Q}L=0,\widetilde{Q}L= 0$: Massless scalar bubble with the same internal masses }
		
		To check the validity of our method, we first consider a scalar bubble with $m_1=m_2=m$ and $q_2^2=0$, defined as
		\begin{align}
			I_{2}^{m_1=m_2,q_2^2=0}=\int {d^D\ell\over i \pi^{D/2}}{1\over (\ell^2-m^2)[(\ell-q_2)^2-m^2]}\co
		\end{align}
		which can be reduced to two tadpoles\footnote{One can check this by using FIRE or direct calculation. }.
		
		We find $Q$  degenerates to a corank-1 matrix
		\begin{align}
			Q&=\left[
			\begin{array}{cc}
				m^2 & m^2 \\
				m^2 & m^2 \\
			\end{array}
			\right]\co~~~\widetilde{Q}=Q^*=\left[
			\begin{array}{cc}
				m^2 & -m^2 \\
				- m^2 & m^2 \\
			\end{array}
			\right]\co
		\end{align}
		and
		\begin{align}
			Q_{(1)}&=Q_{(2)}=m^2,~~
			\adtr{H_1V}=-m^2s_{01},~~\adtr{H_2V}=m^2s_{01},~~\adtr{VV}=m^2 s_{01}^2\ed
		\end{align}
		Notice  that $\adtr{LL}=0, \adtr{VL}=0$, from \eref{deLLdeVL}, we have
		\begin{align}
			I_2
			&={\adtr{H_bV}\over \wt{B}_{2}^{(VV)}}\left[\overtr{LL}_{(b)}I^{(1)}_{2;\what{b}}-{2\overtr{VL}_{(b)}\over D-1}I_{2;\what{b}}\right]
			={D-2\over 2m^2}I_1[m]
		\end{align}
		where $I_1[m]$ means the scalar tadpole with mass $m$ and we have used
		\begin{align}
		\wt{B}_{2}^{(VV)}={-4\adtr{VV}\over D-1},I_{2;\widehat{1}}^{(1)}=2s_{01}I_1[m],~~I_{2;\widehat{1}}^{(2)}=0\ed
		\end{align}
		Then we consider the rank-1 case
		\begin{align}
			I_2^{(1)}={\adtr{H_bV}\over \wt{B}_{3}^{(VV)}}\left[\overtr{LL}_{(b)}I^{(2)}_{n;\what{b}}-{4\overtr{VL}_{(b)}\over D}I^{(1)}_{n;\what{b}}\right]={(D-2)s_{01}\over 2m^2}I_1[m]
		\end{align}
		where we have used
		\begin{align}
		\wt{B}_3^{(VV)}={-8\adtr{VV}\over D},~I_{2;\what{1}}^{(2)}={4(Ds_{01}^2+m^2s_{00})\over D}I_1[m],~~I_{2;\what{2}}^{(2)}={4m^2s_{00}\over D}I_1[m]\ed
		\end{align}
		\item \textbf{$L\widetilde{Q}L\not =0$: Massless scalar bubble with different internal masses }
		
		We then consider the bubbles with degenerate $Q$ and $m_1\not =m_2$.
		Here we will show that it can be reduced to two tadpoles. The equation $\det Q=0$ gives two solutions
		\bea
		q_2^2=(m_1\pm m_2)^2\co~~~\label{4-25}
		\eea
		Here we take the solution $q_2^2=(m_1+m_2)^2$ as an example, and we set
		\bea
		\widetilde{Q}=Q^*=\left[
		\begin{array}{cc}
			m_2^2 & m_1 m_2 \\
		    m_1 m_2 & m_1^2 \\
		\end{array}
		\right]\ed
		\eea
		Notice $\adtr{LL}\equiv L\widetilde{Q}L\not =0$,  we use the formula \eref{deLLnot} which is valid for $r\ge 0$. The reduction for $r=0$ is the same as what we do in \cite{Feng:2022rwj}, which gives
		\begin{align}
			I_{2}=\frac{D-2}{2 (D-3) m_2 \left(m_1+m_2\right)}I_{2;\what{1}}+\frac{D-2}{2 (D-3) m_1 \left(m_1+m_2\right)}I_{2;\what{2}}\ed
		\end{align}
		The reduction for $r\ge 1$ is
		\begin{align}
			I_{2}^{(r)}&={1\over \adtr{LL}}\left[\wt{A}_rI_{2}^{(r-1)}+\wt{B}_r^{(VV)}I_{2}^{(r-2)}+\wt{A}_{r;\what{b}}I^{(r-1)}_{2;\what{b}}+\wt{B}_{r;\what{b}}I^{(r-2)}_{2;\what{b}}\right]\ed \label{bub-de-nonLL}
		\end{align}
		With the expressions below
		\begin{align}
			\adtr{VL}&=m_1 \left(m_1+m_2\right) (R\cdot q_2),~\adtr{H_1V}=m_1 m_2 (R\cdot q_2),~\adtr{H_2V}=m_1^2(R\cdot q_2),\nn
			\adtr{H_1L}&=m_2 \left(m_1+m_2\right),~\adtr{H_2L}=m_1 \left(m_1+m_2\right)\co
		\end{align}
		it is easy to figure out the reduction result
		\begin{align}
			I_2^{(1)}=\frac{\left((D-2) m_1+(D-3) m_2\right) R\cdot q_2}{(D-3) m_2 \left(m_1+m_2\right){}^2}I_{2;\what{1}}+\frac{ R\cdot q_2}{(D-3)  \left(m_1+m_2\right){}^2}I_{2;\what{2}}\ed
		\end{align}
		One can calculate reduction coefficients up to any rank by applying \eref{bub-de-nonLL} iteratively.
	\end{itemize}
   \subsubsection{Triangle}
   We have discussed triangles in the last section, similarly, here to make our expression simple, we can take $\{m_1^2,m_2^2,m_3^2\}=\{1/2,1/3,5/338\},s_{11}=5/7,s_{12}=7/13,s_{22}=3552/5915$. One can check $\det Q=0$, and $LQ^*L=1151/8281\not= 0$. Then we take $\wt{Q}=Q^*$. The reduction relation \eref{deLLnot} becomes
   \begin{align}
   	I_{3}^{(r)}&={1\over \adtr{LL}}\left[\wt{A}_rI_{3}^{(r-1)}+\wt{B}_r^{(VV)}I_{3}^{(r-2)}+\wt{A}_{r;\what{b}}I^{(r-1)}_{3;\what{b}}+\wt{B}_{r;\what{b}}I^{(r-2)}_{3;\what{b}}\right]\ed 
   \end{align}
   where the coefficients are
   \begin{align}
   	\adtr{LL}&={1151\over 8281}\co~~\wt{A}_r={2(D+2r-5)\over D+r-4}\left(-\frac{1151 \left(12 s_{01}-65 s_{02}\right)}{496860}\right)\co\nn
   	\wt{B}_r^{(VV)}&={-4(r-1)\over D+r-4}\frac{1151 \left(12 s_{01}-65 s_{02}\right){}^2}{29811600}\co\nn
   	\wt{A}_{r;\what{b}}&=\left\{\frac{367 s_{01}}{5915}+\frac{16}{91} s_{02},\frac{7}{13} s_{02}-\frac{3552 s_{01}}{5915},\frac{7}{13} s_{01}-\frac{5}{7} s_{02}\right\}\co\nn
   	\wt{B}_{r;\what{b}}&={2(r-1)\over D+r-4}\bigg\{\frac{-4092 s_{01}^2+6643 s_{02} s_{01}-19435 s_{02}^2+1151 s_{00}}{70980},\nn
   	&\newline -\frac{37}{84} s_{02}^2+\frac{19}{35} s_{01} s_{02}-\frac{1151 s_{00}}{41405},-\frac{19}{35} s_{01}^2+\frac{37}{84} s_{02} s_{01}+\frac{1151 s_{00}}{7644}\bigg\}\ed
   \end{align}
   We first solve the reduction of $r=0$ by direct expansion, then applying the recursion relation to obtain 
   \begin{align}
   	I_{3}&={21\over 1151(D-4)} \bigg [4225 (D-2) I_{3;\what{12}}-780 (D-2) I_{3;\what{13}}+455 (D-2) I_{3;\what{23}}\nn&\newline+402 (D-3) I_{3;\what{1}} -592 (D-3)
   	I_{3;\what{2}}
   	+130 (D-3) I_{3;\what{3}}\bigg]\co\nn
   	I_{3}^{(1)}&=\frac{7}{11510 (D-4)} \Bigg[-4225 (D-2) \left(12 s_{01}-65 s_{02}\right)I_{3;\what{12}} +780 (D-2) \left(12 s_{01}-65
   		s_{0,v,2}\right) I_{3;\what{13}}\nn
   		&\newline -455 (D-2) \left(12 s_{01}-65 s_{02}\right) I_{3;\what{23}}+130
   		 \left((37 D-160) s_{01}+65 s_{02}\right)I_{3;\what{3}}\nn
   		&\newline +2  \left(65 (692-247 D) s_{02}+3552
   		s_{0,v,1}\right)I_{3;\what{2}}-2  \left((2045 D-5768) s_{01}+65 (667-217 D) s_{02}\right)I_{3;\what{1}}\Bigg]\ed
     \end{align}
 One can check it is right for higher ranks, for example,
 \begin{align}
 	C^{(2)}_{3\to 3;\what{1}}&=-\frac{7}{209256979500 (D-4) (D-2) (D-1)} \bigg[\nn
 	&\newline \left(8552241125 D^3-76543388850 D^2+225731680480 D-141381541152\right) s_{01}^3\nn
 	&\newline -195 \left(907499425 D^3-6547444365 D^2+16045670426 D-10487335800\right) s_{02} s_{01}^2\nn
 	&\newline +12675 \left(96297005 D^3-609524352 D^2+1222062838 D-714186768\right) s_{02}^2 s_{01}\nn
 	&\newline -274625 \left(10218313 D^3-64319451 D^2+123317324 D-69406506\right) s_{02}^3\nn 
 	&
 	\newline +3470265 (D-4) s_{00} \left((2045 D-1678) s_{01}+65 (233-217 D) s_{02}\right)\bigg]\ed
 \end{align}
    
	%%%%%%%%%%%%%%%%%%%%%%%%
	
	\section{Conclusion and Outlook}
	\label{sec:discussion}
	In this paper, we derive non-trivial recursive relations for reducing one-loop tensor integrals and extend it for degenerate $Q$. As in \cite{Feng:2022rwj}, we first express one-loop integrals as a sum of integrals $E_{n,k}[V^i]$ in projective space. However, unlike the formal approach where each $E_{n,k}[V^i]$ is naively reduced to a scalar basis, we use the basic recursive relation of $E_{n,k}[V^i]$ to derive a single recursion relation for one-loop Feynman integrals. We start by discussing the non-degenerated case and demonstrate that a $r$-rank tensor integral can be reduced to $(r - 1)$ and $(r - 2)$-rank tensor integrals of the same sector and the sub-sector (with one propagator removed). This recursive relation at the integral level avoids the need for an $E_{n,k}[V^i]$ expansion and term-by-term reduction, Furthermore, the previous computation results can be reused in the next iteration, which significantly speeds up the calculation process.  We present the results of the time taken for both methods in Table \underline{\ref{running time}} and Table \underline{\ref{running time-2}}. Additionally, the recursive relation enables us to study the general structure, such as singularity, of reduction coefficients, as shown by our explicit examples in the main text and the results listed in Appendix \ref{appdix:highernr}.
	
	\begin{figure}[h]
		\begin{center}
			\includegraphics[scale=1]{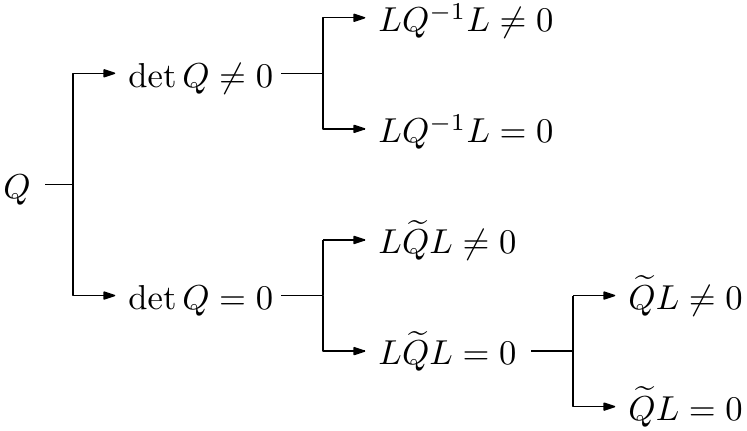}
			\caption{Reduction relations classification}
			\label{fig:reduction}
		\end{center}
	\end{figure}

    Then, we present an explicit derivation for the recursion relations for degenerate cases by modifying the original recursions for non-degenerate $Q$. Although it may appear to be an ad-hoc patch, this approach actually provides a coherent framework with an explicit derivation given in Appendix \ref{appdix:lowerterm}, where one can replace $Q^{-1}$ with a general symmetric matrix $\widetilde{Q}$. The reduction framework is shown in Figure \underline{\ref{fig:reduction}}. Depending on the determinant of $Q$ and the choice of $\widetilde{Q}$, we utilize one of five different recursions. The recursion relations are expressed in lower rank or lower topology integrals with the \textit{same} dimensions, except in the very specific cases where both $Q$ and $Q_{(b)}$ are degenerate. The reduction coefficient tensor structure and runtime analysis for degenerate cases are similar to those for non-degenerate cases. Therefore, this article will not delve into further detail on this topic.
    
	There are some things to be explored in future, which we briefly comment below.
	\begin{itemize}
		\item  
		The occurrence in which both $Q$ and $Q_{(b)}$ are degenerate can be resolved by introducing additional $\wt{Q_b}$ in the reduction of $E_{n-1,k}$. This process is similar to the approach taken in the present context, albeit more intricate.
		\item 
		As demonstrated in \cite{Feng:2022rfz}, degenerate integrals can also be reduced by means of taking limits based on the expressions of reduction coefficients pertaining to non-degenerate $Q$. In this context, an anzatz is proposed whereby one basis splits into a combination of other bases. The unknown coefficients within the anzatz are then evaluated by requiring that all singularities cancel each other out.
		\item 
		In the case of one-loop integrals, $E_{n,k}$ is defined by only two tensor structures, namely $V_IX^I$ and $L_IX^I$, enabling the extraction of all $V$ elements in order to reduce it. However, in attempting to extend this same methodology to two-loop integrals, the tensor integrals must be written in projective space, revealing more complex tensor structures such as $W_{IJ}X^{I}{J}$, $V_IX^I$, $(H_b)_IX^I$, and so forth, coupled with a cubic polynomial, $Q_{IJK}X^{I}X^JX^K$, appearing in the denominator. Due to the difficulty in calculating the inverse of $Q$ and extracting the unwanted $V$ elements, a different approach must be taken. I believe this is related to the appearance of irreducible scalar product in the case of two loops, but further research is needed to confirm this.
	\end{itemize}

	\section*{Acknowledgments}
	% % % % % % % % % %
	I would like to thank Bo Feng for the inspiring discussion and guidance. This work is supported by Chinese NSF funding under Grant No.11935013, No.11947301, No.12047502 (Peng Huanwu Center).

	\appendix
	%%%%%%%%%%%%%%%%%%
	\section{Expression of \textit{lower terms}}
\label{appdix:lowerterm}

As we have stated in the main text, the recursion relations \eref{lower-d-2} and \eref{lower-samed} contain two forms of lower-topology terms. The first form is simpler formally, but includes terms of various dimensions. In contrast, the second form does not involve dimension shifting integrals, making it a preferred calculation option except for degenerate cases of $Q_{(b)}$. In this article, we derive both forms of the lower-topology terms utilizing the basic recursion relations of $E_{n,k}$, operating under the assumption that $Q$ and its submatrices are non-degenerate. For scenarios where $Q$ is degenerate, we suggest replacing $Q^{-1}$ with an arbitrary symmetric matrix $\widetilde{Q}$ during calculations.

We derive \eref{lower-d-2} first  by direct expansion:
\begin{align}
	{\lt{n}{r}\over \overtr{LL}}&=I_{n}^{(r)}-{1\over \overtr{LL}}\left[A_rI_{n}^{(r-1)}+B_rI_{n}^{(r-2)}\right]\nn
	&={\Gamma(n-D/2-r)\over (-)^{n+r}}\sum_{i=0}^{r}\mathscr{C}^{D/2+r-n}_{r,i} 
	(R^2)^{r-i\over 2} E_{n,n-D-r+i}[V^i]-a_r{\overtr{VL}\over  \overtr{LL}} \nn
	&\newline\times{\Gamma(n-D/2-r+1)\over (-)^{n+r-1}}\sum_{i=0}^{r-1}\mathscr{C}^{D/2+r-1-n}_{r-1,i} 
	(R^2)^{r-1-i\over 2} E_{n,n-D-r+1+i}[V^i]-b_r{R^2-\overtr{VV}\over  \overtr{LL}} \nn
	&\newline \times{\Gamma(n-D/2-r+2)\over (-)^{n+r-2}}\sum_{i=0}^{r-2}\mathscr{C}^{D/2+r-2-n}_{r-2,i} 
	(R^2)^{r-2-i\over 2} E_{n,n-D-r+2+i}[V^i]~~ \label{RecurExpand}
\end{align} 
where we have defined
\begin{align}
	a_r&={2(D+2r-n-2)\over D+r-n-1},~~
	b_r={4(r-1)\over  D+r-n-1}\ed
\end{align}
Since the only integrals present in the lower terms are of lower typologies, they must be expressed in terms of $E_{n,k}^{(b)}$. The proof strategy involves reducing the three summations of $E_{n,k}$ according to the \eref{general-resur} principle. We aim for the top sector terms, specifically, $E_{n,k}$, to cancel each other out and leave behind only the $E_{n,k}^{(b)}$ terms. It is worth noting that only the first summation on the right-hand side (RHS) does not contain the denominator ${\overtr{LL}}$. Therefore, a recursion is required to find a reduction for $E_{n,k}$ to a fraction containing the denominator $\overtr{LL}$. Two different recursion relations can be considered for different tensor orders.
\begin{itemize}
	\item $T=V\otimes V^i\otimes L^{k-i-1}$
	\begin{align}
		E_{n,k}[V^{i+1}]
		&=\alpha_{n,k}\Big [{i }\overtr{VV}E_{n,k-2}[V^{i-1}]+(k-i-1)\overtr{VL}E_{n,k-2}[V^i]\nn
		&\newline+\overtr{H_bV}E^{(b)}_{n-1,k-1}[V^{i}]\Big]\co \label{V-first-recur}
	\end{align}
	\item $T=L\otimes V^{i+1}\otimes L^{k-i-2}$
	\begin{align}
		E_{n,k}[V^{i+1}]
		&=\alpha_{n,k}\Big [{(i+1) }\overtr{VL}E_{n,k-2}[V^{i}]+(k-i-2)\overtr{LL}E_{n,k-2}[V^{i+1}]\nn
		&\newline+\overtr{H_bL}E^{(b)}_{n-1,k-1}[V^{i+1}]\Big]\ed
	\end{align}
\end{itemize}
Rearrange the second relation we get 
\begin{align}
	E_{n,k}[V^{i}]={E_{n,k+2}[V^{i}]/\a_{n,k+2}-i\overtr{VL}E_{n,k}[V^{i-1}]-\overtr{H_bL}E^{(b)}_{n-1,k+1}[V^{i}]\over (k+1-i)\overtr{LL}}\ed\label{L-first-recur}
\end{align}
The above formula applies for $i > 0$, but it can also be extended to the case where $i = 0$ by replacing $V$ with $L$:
\begin{align}
	E_{n,k}={E_{n,k+2}/\a_{n,k+2}-i\overtr{LL}E_{n,k}-\overtr{H_bL}E^{(b)}_{n-1,k+1}\over (k+1-i)\overtr{LL}}\co
\end{align}
then it becomes safe to take $i=0$:
\begin{align}
	E_{n,k}={E_{n,k+2}/\a_{n,k+2}-\overtr{H_bL}E^{(b)}_{n-1,k+1}\over (k+1)\overtr{LL}}\ed\label{LV0-first-recur}
\end{align}
For the sake of simplicity, we introduce the notation $k_0=n-D-r$, $k(i)=k_0+i$, and $s=D/2+r-n$. Using equations \eref{L-first-recur} and \eref{LV0-first-recur}, we can reduce the first summation on the right-hand side of equation \eref{RecurExpand}:
\begin{align}
	&{\Gamma(-s)\over (-)^{n+r}}\sum_{i=0}^{r}\mathscr{C}^{s}_{r,i} 
	(R^2)^{r-i\over 2} E_{n,k(i)}[V^i]\nn
	=&{\Gamma(-s)\over (-)^{n+r}}\mathscr{C}^{s}_{r,0} 
	(R^2)^{r\over 2}{1\over k(0)+1}{E_{n,k_0+2}/\a_{n,k_0+2}-\overtr{H_bL}E^{(b)}_{n-1,k_0+1}\over \overtr{LL}}\nn
	&+{\Gamma(-s)\over (-)^{n+r}}\sum_{i>0}^{r}\mathscr{C}^{s}_{r,i} 
	(R^2)^{r-i\over 2} {\boxed{E_{n,k(i)+2}[V^{i}]}/\a_{n,k(i)+2}-i\overtr{VL}E_{n,k(i)}[V^{i-1}]-\overtr{H_bL}E^{(b)}_{n-1,k(i)+1}[V^{i}]\over (k+1-i)\overtr{LL}} \label{Sum1-exp}\ed
\end{align}
To simplify the equation, we utilize equation \eref{V-first-recur} to reduce the boxed term in \eref{Sum1-exp}. In this process, we box the updated portion for emphasis.
\begin{align}
	&{{\Gamma(-s)}\mathscr{C}^{s}_{r,0} 
		(R^2)^{r\over 2}\over (-)^{n+r} (k(0)+1)\overtr{LL}}\Big[E_{n,k_0+2}/\a_{n,k_0+2}-\overtr{H_bL}E^{(b)}_{n-1,k_0+1}\Big]+{\Gamma(-s)\over (-)^{n+r}}\sum_{i>0}^{r} {\mathscr{C}^{s}_{r,i} 
		(R^2)^{r-i\over 2}\over (k+1-i)\overtr{LL}}\nn
	&\newline \times \Big [\boxed{(i-1) \overtr{VV}E_{n,k(i)}[V^{i-2}]+(k(i)+2-i)\overtr{VL}E_{n,k-2}[V^i]+\overtr{H_bV}E^{(b)}_{n-1,k(i)+1}[V^{i-1}]}\nn
	& \newline -i\overtr{VL}E_{n,k(i)}[V^{i-1}]-\overtr{H_bL}E^{(b)}_{n-1,k(i)+1}[V^{i}]\Big]\ed
\end{align}
Now, we try to deal with other two summations in \eref{RecurExpand}:
\begin{align}
	&-a_r{\overtr{VL}\over  \overtr{LL}}
	{\Gamma(-s+1)\over (-)^{n+r-1}}\sum_{i=0}^{r-1}\mathscr{C}^{s-1}_{r-1,i} 
	(R^2)^{r-1-i\over 2} E_{n,k(i)+1}[V^i]\nn
	&-b_r{\boxed{R^2-\overtr{VV}}\over  \overtr{LL}} 
	{\Gamma(-s+2)\over (-)^{n+r-2}} \sum_{i=0}^{r-2}\mathscr{C}^{s-2}_{r-2,i} 
	(R^2)^{r-2-i\over 2} E_{n,k(i)+2}[V^i]\ed \label{other-two-sums}
\end{align}
To clarify, the terms in the box frame, denoted as $R^2-\overtr{VV}$, allow us to split the second line into two parts. We can absorb the $R^2$ term into the summation, resulting in the modification of \eref{other-two-sums}.
\begin{align}
  &-a_r{\overtr{VL}\over  \overtr{LL}}
	{\Gamma(-s+1)\over (-)^{n+r-1}}\sum_{i=0}^{r-1}\mathscr{C}^{s-1}_{r-1,i} 
	(R^2)^{r-1-i\over 2} E_{n,k(i)+1}[V^i]-b_r{-\overtr{VV}\over  \overtr{LL}} 
	{\Gamma(-s+2)\over (-)^{n+r-2}}\nn
	&\newline \times \sum_{i=0}^{r-2}\mathscr{C}^{s-2}_{r-2,i} 
	(R^2)^{r-2-i\over 2} E_{n,k(i)+2}[V^i]-b_r{1\over  \overtr{LL}} 
	{\Gamma(-s+2)\over (-)^{n+r-2}}\sum_{i=0}^{r-2}\mathscr{C}^{s-2}_{r-2,i} 
	(R^2)^{r-i\over 2} E_{n,k(i)+2}[V^i]\co \label{other-two-sums-split}
\end{align}
then we reduce the last part in \eref{other-two-sums-split} using \eref{V-first-recur}
\begin{align}
	&\sum_{i=0}^{r-2}\mathscr{C}^{s-2}_{r-2,i} 
	(R^2)^{r-i\over 2} E_{n,k(i)+2}[V^i]=\mathscr{C}^{s-2}_{r-2,0} 
	(R^2)^{r\over 2} E_{n,k_0+2}[V^0]+\sum_{i>0}^{r-2}\mathscr{C}^{s-2}_{r-2,i} 
	(R^2)^{r-i\over 2}  \alpha_{n,k(i)+2}\nn
	&\times \Big [{(i-1) }\overtr{VV}E_{n,k(i)}[V^{i-2}]+(k(i)+2-i)\overtr{VL}E_{n,k(i)}[V^{i-1}]+\overtr{H_bV}E^{(b)}_{n-1,k(i)+1}[V^{i-1}]\Big]\ed\nonumber
\end{align}
Sum over all terms, we find RHS of \eref{RecurExpand} becomes 
\allowdisplaybreaks
\begin{align}
	&{{\Gamma(-s)}\mathscr{C}^{s}_{r,0} 
		(R^2)^{r\over 2}\over (-)^{n+r} (k(0)+1)\overtr{LL}}\Big[E_{n,k_0+2}/\a_{n,k_0+2}-\overtr{H_bL}E^{(b)}_{n-1,k_0+1}\Big]\nn
	+&{\Gamma(-s)\over (-)^{n+r}}\sum_{i>0}^{r} {\mathscr{C}^{s}_{r,i} 
		(R^2)^{r-i\over 2}\over (k+1-i)\overtr{LL}} \Big [{(i-1) }\overtr{VV}E_{n,k(i)}[V^{i-2}]+(k(i)+2-i)\overtr{VL}E_{n,k-2}[V^i]\nn
	&+\overtr{H_bV}E^{(b)}_{n-1,k(i)+1}[V^{i-1}]-i\overtr{VL}E_{n,k(i)}[V^{i-1}]-\overtr{H_bL}E^{(b)}_{n-1,k(i)+1}[V^{i}]\Big]\nn
	-&a_r{\overtr{VL}\over  \overtr{LL}}
	{\Gamma(-s+1)\over (-)^{n+r-1}}\sum_{i=0}^{r-1}\mathscr{C}^{s-1}_{r-1,i} 
	(R^2)^{r-1-i\over 2} E_{n,k(i)+1}[V^i]\nn
	+&b_r{\overtr{VV}\over  \overtr{LL}} 
	{\Gamma(-s+2)\over (-)^{n+r-2}}\sum_{i=0}^{r-2}\mathscr{C}^{s-2}_{r-2,i} 
	(R^2)^{r-2-i\over 2} E_{n,k(i)+2}[V^i]
	-{b_r\over  \overtr{LL}} 
	{\Gamma(-s+2)\over (-)^{n+r-2}}\mathscr{C}^{s-2}_{r-2,0} 
	(R^2)^{r\over 2} E_{n,k_0+2}[V^0]\nn
	-&{b_r\over  \overtr{LL}} 
	{\Gamma(-s+2)\over (-)^{n+r-2}}\sum_{i>0}^{r-2}\mathscr{C}^{s-2}_{r-2,i} 
	(R^2)^{r-i\over 2}  \alpha_{n,k(i)+2}\Big [{(i-1) }\overtr{VV}E_{n,k(i)}[V^{i-2}]\nn
	&\newline +(k(i)+2-i)\overtr{VL}E_{n,k(i)}[V^{i-1}]+\overtr{H_bV}E^{(b)}_{n-1,k(i)+1}[V^{i-1}]\Big]\ed
\end{align}
By utilizing the definition of ${\mathscr C}, a_r,$ and $b_r$, it has been determined that the contribution of $E_{n,k}$ vanishes. Then the lower terms become
\begin{align}
	\lt{n}{r}&={\Gamma(-s)\over (-)^{n+r}}\sum_{i>0}^{r}\mathscr{C}^{s}_{r,i} 
	(R^2)^{r-i\over 2} {\overtr{H_bV}E^{(b)}_{n-1,k(i)+1}[V^{i-1}]-\overtr{H_bL}E^{(b)}_{n-1,k(i)+1}[V^{i}]\over (k+1-i)}\nn
	&\newline +{\Gamma(-s)\over (-)^{n+r}}\mathscr{C}^{s}_{r,0} 
	(R^2)^{r\over 2}{-\overtr{H_bL}E^{(b)}_{n-1,k_0+1}\over k_0+1}\nn
	&\newline -b_r{\Gamma(-s+2)\over (-)^{n+r-2}}\sum_{i>0}^{r-2}\mathscr{C}^{s-2}_{r-2,i} 
	(R^2)^{r-i\over 2} \a_{n,k(i)+2}\overtr{H_bV}E^{(b)}_{n-1,k(i)+1}[V^{i-1}]\nn
	&={\Gamma(-s) (-)^{n-1+r}\overtr{H_bL}\over(n-r-D+1)}\sum_{i=0}^{r}\mathscr{C}^{s}_{r,i} 
	(R^2)^{r-i\over 2} {E^{(b)}_{n-1,k(i)+1}[V^{i}]}\nn
	&\newline+{\Gamma(-s)(-)^{n+r}\overtr{H_bV}\over (n-r-D+1)}\sum_{i=0}^{r-1}\mathscr{C}^{s}_{r,i+1} 
	(R^2)^{r-i-1\over 2} {E^{(b)}_{n-1,k(i)+2}[V^{i}]}\nn
	&\newline+{4(r-1)(-)^{n+r}\Gamma(-s+2)\overtr{H_bV}\over n-r-D+1}\sum_{i=0}^{r-3}\mathscr{C}^{s-2}_{r-2,i+1} 
	(R^2)^{r-i-1\over 2} \a_{n,k(i)+3}E^{(b)}_{n-1,k(i)+2}[V^{i}]\ed
\end{align}
After some algebra, one reaches
\bea
\lt{n}{r}={\overtr{H_bL}I^{(r)}_{n;D-2,\what{b}}+ 2\overtr{H_bV}I^{(r-1)}_{n;D-2,\what{b}} \over D+r-n-1}\label{A1-lower-d-2}\ed
\eea
Then we establish the validity of \eref{lower-samed} through the use of \eref{V-first-recur} to simplify \eref{A1-lower-d-2}.
\begin{align}
	&\overtr{H_bL}I^{(r)}_{n;D-2,\what{b}}+ 2\overtr{H_bV}I^{(r-1)}_{n;D-2,\what{b}}\nn
	=&\overtr{H_bL}{\Gamma(-s)\over (-)^{n-1+r}}\sum_{i=0}^{r}\mathscr{C}^{s}_{r,i} 
	(R^2)^{r-i\over 2} E^{(b)}_{n-1,k(i)+1}[V^i]+2\overtr{H_bV}{\Gamma(-s+1)\over (-)^{n+r}}\sum_{i=0}^{r-1}\mathscr{C}^{s-1}_{r-1,i} 
	(R^2)^{r-1-i\over 2} E^{(b)}_{n-1,k(i)+2}[V^i]\nn
	=&\overtr{H_bL}{\Gamma(-s)\over (-)^{n-1+r}}\mathscr{C}^{s}_{r,0} 
	(R^2)^{r\over 2} E^{(b)}_{n-1,k_0+1}+2\overtr{H_bV}{\Gamma(-s+1)\over (-)^{n+r}}\mathscr{C}^{s-1}_{r-1,0} 
	(R^2)^{r-1\over 2} E^{(b)}_{n-1,k(i)+2}\nn
	&+\overtr{H_bL}{\Gamma(-s)\over (-)^{n-1+r}}\sum_{i>0}^{r}\mathscr{C}^{s}_{r,i} 
	(R^2)^{r-i\over 2}\alpha_{n-1,k(i)+1}\Big [{(i-1) }\overtr{VV}_{(b)}E^{(b)}_{n-1,k(i)-1}[V^{i-2}]\nn
	&+(k+1-i)\overtr{VL}_{(b)}E^{(b)}_{n-1,k(i)-1}[V^{i-1}]+\overtr{H_cV}_{(b)}E^{(bc)}_{n-2,k(i)}[V^{i-1}]\Big]\nn
	&+2\overtr{H_bV}{\Gamma(-s+1)\over (-)^{n+r}}\sum_{i>0}^{r-1}\mathscr{C}^{s-1}_{r-1,i} 
	(R^2)^{r-1-i\over 2} \alpha_{n-1,k(i)+2}\Big [{i }\overtr{VL}_{(b)}E^{(b)}_{n-1,k(i)-1}[V^{i-2}]\nn
	&+(k+1-i)\overtr{VL}_{(b)}E^{(b)}_{n-1,k(i)-1}[V^{i-1}]+\overtr{H_cL}_{(b)}E^{(bc)}_{n-2,k(i)+1}[V^{i}]\Big]\ed
\end{align}
The contribution of $E^{(bc)}$ can be verified to vanish through verification of the corresponding identity.
\bea
\overtr{H_bL}\overtr{H_cV}_{(b)}+\overtr{H_cL}\overtr{H_bV}_{(c)}= \overtr{H_bV}\overtr{H_cL}_{(b)}+\overtr{H_cV}\overtr{H_bL}_{(c)}\ed
\eea
After performing some algebraic manipulations, it has been determined that the remaining items are
\begin{align}
	\lt{n}{r}&=\left[\overtr{H_bL}\overtr{VL}_{(b)}-\overtr{H_bV}\overtr{LL}_{(b)}\right]I^{(r-1)}_{n;\what{b}}\nn
	&\newline+{2(r-1)\left[\overtr{H_bL}R^2+\overtr{H_bV}\overtr{VL}_{(b)}-\overtr{H_bL}\overtr{VV}_{(b)}\right]\over D+r-n-1 }I_{n;\what{b}}^{(r-2)}\ed
\end{align}
\section{Reduction results for higher $n,r$}
\label{appdix:highernr}
Here we provide a list of reduction results for $n=3,4,5$ and $r=6$ based on the divergence of $\overtr{LL}$. While the lower order divergent parts may be too complicated to display, we present the complete results for the top sector and the results up to sub-leading divergence in sub-sectors. To avoid unnecessary complexity, we set $D=4$ for $n=3,4$ and $D=6$ for $n=5$. Additionally, we define $A_{\{i\}}=A_{(\{i\}^c)}$, where $\{i\}$ is the label list of the propagators survived. In some cases, to simplify expressions, we use $x_{\{i\}}={1\over \overtr{LL}_{\{i\}}}$.
\subsection{Reduction coefficients for triangle in $D=4$}
In the main text, we have provided some results. In this section, we present the sub-leading divergence of the tadpole coefficient, namely $C_{3\to 3;\what{23}}^{(6)}\Big\vert_{\overtr{LL}^{-4}}$.
\begin{align}
	&\frac{616 \overtr{LV}_{\{1,2\}}^2 \overtr{LH_3} \left(\overtr{LV}_{\{1\}} \overtr{LH_2}_{\{1,2\}}-\overtr{LL}_{\{1\}} \overtr{VH_2}_{\{1,2\}}\right) \overtr{LV}^3}{5 \overtr{LL}_{\{1,2\}}^2}\nn
	&+\frac{8\overtr{LV}^2}{3 \overtr{LL}_{\{1,3\}}} \Bigg(\frac{693 \overtr{LV}_{\{1,3\}} \overtr{LV} \overtr{LH_2} \overtr{LH_3}_{\{1,3\}} \overtr{LV}_{\{1\}}^2}{20 \overtr{LL}_{\{1\}}}\nn
	&-\frac{231}{5} \overtr{LV}_{\{1,3\}} \overtr{LV} \overtr{LH_3}_{\{1,3\}} \overtr{VH_2} \overtr{LV}_{\{1\}}+\frac{231}{5} \overtr{LL}_{\{1\}} \overtr{LV}_{\{1,3\}} \overtr{LV} \overtr{VH_2} \overtr{VH_3}_{\{1,3\}}\nn&+\frac{441}{10} \overtr{LV}_{\{1,3\}} \overtr{LH_2} \left(R^2-\overtr{VV}\right) \left(\overtr{LV}_{\{1\}} \overtr{LH_3}_{\{1,3\}}-\overtr{LL}_{\{1\}} \overtr{VH_3}_{\{1,3\}}\right)\nn&+\frac{231}{10} \overtr{LV} \left(\overtr{LH_2} \left(R^2-\overtr{VV}_{\{1,3\}}\right)+\overtr{LV}_{\{1,3\}} \overtr{VH_2}\right) \left(\overtr{LV}_{\{1\}} \overtr{LH_3}_{\{1,3\}}-\overtr{LL}_{\{1\}} \overtr{VH_3}_{\{1,3\}}\right)\nn&-\frac{231}{20} \overtr{LV}_{\{1,3\}} \overtr{LV} \overtr{LH_2} \left(\overtr{LH_3}_{\{1,3\}} \left(\overtr{VV}_{\{1\}}-R^2\right)+2 \overtr{LV}_{\{1\}} \overtr{VH_3}_{\{1,3\}}\right)\Bigg) \nn&-\frac{2\overtr{LV}^2 }{45 \overtr{LL}_{\{1\}}}\Bigg(2079 \overtr{LV} \overtr{LH_3}_{\{1,3\}} \overtr{VH_2} \overtr{LV}_{\{1\}}^2\Bigg) +(2\leftrightarrow 3)\ed
\end{align}
\subsection{Reduction coefficients for box in $D=4$}
\textbf{Box coefficient}:
\begin{align}
	\frac{2048 \overtr{LV}^6}{\overtr{LL}^6}+\frac{3072 \overtr{LV}^4 \left(R^2-\overtr{VV}\right)}{\overtr{LL}^5} +\frac{1152 \overtr{LV}^2 \left(R^2-\overtr{VV}\right){}^2}{\overtr{LL}^4}+\frac{64 \left(R^2-\overtr{VV}\right){}^3}{\overtr{LL}^3}\ed
\end{align}
\textbf{Triangle coefficient}:

\textsf{Leading divergence} $C_{4\to 4;\what{4}}^{(6)}\Big\vert_{\overtr{LL}^{-6}}$:
\begin{align}
	&1024 \overtr{LH_4} \overtr{LV}_{\{1,2,3\}} \overtr{LV}^5-1024 \overtr{LL}_{\{1,2,3\}} \overtr{VH_4} \overtr{LV}^5\ed
\end{align}

\textsf{Sub-leading divergence} $C_{4\to 4;\what{4}}^{(6)}\Big\vert_{\overtr{LL}^{-5}}$:
\begin{align}
	&512 R^2 \overtr{LH_4} \overtr{LV}^4+1024 R^2 \overtr{LH_4} \overtr{LV}_{\{1,2,3\}} \overtr{LV}^3-1024 R^2 \overtr{LL}_{\{1,2,3\}} \overtr{VH_4} \overtr{LV}^3\nn
	&+\frac{512 \overtr{LH_4} \overtr{LV}_{\{1,2,3\}}^2 \overtr{LV}^4}{\overtr{LL}_{\{1,2,3\}}}+1024 \overtr{LL}_{\{1,2,3\}} \overtr{VV} \overtr{VH_4} \overtr{LV}^3\nn
	&-1024 \overtr{VV} \overtr{LH_4} \overtr{LV}_{\{1,2,3\}} \overtr{LV}^3-512 \overtr{VV}_{\{1,2,3\}} \overtr{LH_4} \overtr{LV}^4\ed
\end{align}
\textbf{Bubble coefficient}:

\textsf{Leading divergence} $C_{4\to 4;\what{34}}^{(6)}\Big\vert_{\overtr{LL}^{-5}}$:
\begin{align}
	&256 \overtr{LV}^4 \Bigg(x_{\{1,2,3\}} \overtr{LH_4} \overtr{LV}_{\{1,2,3\}} \left(\overtr{LH_3}_{\{1,2,3\}} \overtr{LV}_{\{1,2\}}-\overtr{LL}_{\{1,2\}} \overtr{VH_3}_{\{1,2,3\}}\right)\nn &-\overtr{LH_4}_{\{1,2,4\}} \overtr{VH_3} \overtr{LV}_{\{1,2\}}+\overtr{LL}_{\{1,2\}} \overtr{VH_3} \overtr{VH_4}_{\{1,2,4\}}\Bigg)+(3\leftrightarrow 4)\ed
\end{align}
\textbf{Tadpole coefficient}:

\textsf{Leading divergence} $C_{4\to 4;\what{234}}^{(6)}\Big\vert_{\overtr{LL}^{-4}}$:
\begin{align}
	&64 \overtr{LV}^3 \Bigg[\left(x_{\{1,2,4\}} \overtr{LH_3} \overtr{LV}_{\{1,2,4\}}-\overtr{VH_3}\right) \Big[\left(\overtr{LV}_{\{1\}} \overtr{LH_4}_{\{1,4\}}-\overtr{LL}_{\{1\}} \overtr{VH_4}_{\{1,4\}}\right)\nn&\newline\times \left(x_{\{1,4\}} \overtr{LH_2}_{\{1,2,4\}} \overtr{LV}_{\{1,4\}}-\overtr{VH_2}_{\{1,2,4\}}\right)\Big]
	\Bigg]+\text{permutations}\ \text{of}\ (2,3,4)\ed
\end{align}
\subsection{Reduction coefficients for pentagon in $D=6$}
\textbf{Pentagon coefficient}:
\begin{align}
	\frac{1260 \overtr{LV}^4 \left(R^2-\overtr{VV}\right)}{\overtr{LL}^5}+\frac{420 \overtr{LV}^2 \left(R^2-\overtr{VV}\right){}^2}{\overtr{LL}^4}+\frac{20 \left(R^2-\overtr{VV}\right){}^3}{\overtr{LL}^3}+\frac{924 \overtr{LV}^6}{\overtr{LL}^6}\ed
\end{align}
\textbf{Box coefficient}:

\textsf{Leading divergence} $C_{5\to 5;\what{5}}^{(6)}\Big\vert_{\overtr{LL}^{-6}}$:
\begin{align}
	462 \overtr{LV}^5 \left(\overtr{LH_5} \overtr{LV}_{\{1,2,3,4\}}-\overtr{LL}_{\{1,2,3,4\}} \overtr{VH_5}\right)\ed
\end{align}

\textsf{Sub-leading divergence} $C_{5\to 5;\what{5}}^{(6)}\Big\vert_{\overtr{LL}^{-5}}$:
\begin{align}
	&-14 \overtr{LV}^3 \Bigg(\overtr{LV}_{\{1,2,3,4\}} \left(11 \overtr{VH_5} \overtr{LV}-34 \overtr{LH_5} \left(R^2-\overtr{VV}\right)\right)
	\nn&\newline\newline-11 \overtr{LH_5} \overtr{LV} \left(R^2-\overtr{VV}_{\{1,2,3,4\}}\right)\Bigg)+308 x_{\{1,2,3,4\}} \overtr{LH_5} \overtr{LV}_{\{1,2,3,4\}}^2 \overtr{LV}^4\nn
	&\newline-\frac{476 \overtr{VH_5} \overtr{LV}^3 \left(R^2-\overtr{VV}\right)}{x_{\{1,2,3,4\}}}\ed
\end{align}
\textbf{Triangle coefficient}:

\textsf{Leading divergence} $C_{5\to 5;\what{45}}^{(6)}\Big\vert_{\overtr{LL}^{-5}}$:
\begin{align}
	&\frac{154 \overtr{LH_5} \overtr{LV}_{\{1,2,3,4\}} \overtr{LV}^4 \left(\overtr{LH_4}_{\{1,2,3,4\}} \overtr{LV}_{\{1,2,3\}}-\overtr{LL}_{\{1,2,3\}} \overtr{VH_4}_{\{1,2,3,4\}}\right)}{\overtr{LL}_{\{1,2,3,4\}}}\nn
	&+154 \overtr{LV}^4 \Bigg(\overtr{LL}_{\{1,2,3\}} \overtr{VH_4} \overtr{VH_5}_{\{1,2,3,5\}}-\overtr{LV}_{\{1,2,3\}} \overtr{LH_5}_{\{1,2,3,5\}} \overtr{VH_4}\Bigg)+(4 \leftrightarrow 5)\ed
\end{align}
%\begin{align}	&\frac{154 \overtr{LH_5} \overtr{LV}_{\{1,2,3,4\}} \overtr{LV}^4 \left(\overtr{LH_4}_{\{1,2,3,4\}} \overtr{LV}_{\{1,2,3\}}-\overtr{LL}_{\{1,2,3\}} \overtr{VH_4}_{\{1,2,3,4\}}\right)}{\overtr{LL}_{\{1,2,3,4\}}}\nn	&+\frac{154 \overtr{LH_4} \overtr{LV}_{\{1,2,3,5\}} \overtr{LV}^4 \left(\overtr{LH_5}_{\{1,2,3,5\}} \overtr{LV}_{\{1,2,3\}}-\overtr{LL}_{\{1,2,3\}} \overtr{VH_5}_{\{1,2,3,5\}}\right)}{\overtr{LL}_{\{1,2,3,5\}}}\nn	&+154 \overtr{LV}^4 \Bigg(\overtr{LL}_{\{1,2,3\}} \left(\overtr{VH_4} \overtr{VH_5}_{\{1,2,3,5\}}+\overtr{VH_4}_{\{1,2,3,4\}} \overtr{VH_5}\right)\nn&\newline-\overtr{LV}_{\{1,2,3\}} \left(\overtr{LH_5}_{\{1,2,3,5\}} \overtr{VH_4}+\overtr{LH_4}_{\{1,2,3,4\}} \overtr{VH_5}\right)\Bigg)\end{align}
\textbf{Bubble coefficient}:

\textsf{Leading divergence} $C_{5\to 5;\what{345}}^{(6)}\Big\vert_{\overtr{LL}^{-4}}$:
\begin{align}
	&\frac{231}{5} \overtr{LV}^3 \left(x_{\{1,2,3,5\}} \overtr{LH_4} \overtr{LV}_{\{1,2,3,5\}}-\overtr{VH_4}\right)\left(x_{\{1,2,5\}} \overtr{LH_3}_{\{1,2,3,5\}} \overtr{LV}_{\{1,2,5\}}-\overtr{VH_3}_{\{1,2,3,5\}}\right)\nn
	&\newline\times \left(\overtr{LH_5}_{\{1,2,5\}} \overtr{LV}_{\{1,2\}}-\overtr{LL}_{\{1,2\}} \overtr{VH_5}_{\{1,2,5\}}\right) + \text{permutations}\ \text{of}\ (3,4,5)\ed
\end{align}
\textbf{Tadpole coefficient}:

\textsf{Leading divergence} $C_{5\to 5;\what{2345}}^{(6)}\Big\vert_{\overtr{LL}^{-3}}$:
\begin{align}
	&\frac{66}{5} \overtr{LV}^2 \Bigg(x_{\{1,2,3,4\}} \overtr{LH_5} \overtr{LV}_{\{1,2,3,4\}} \left(\overtr{LV}_{\{1\}} \overtr{LH_4}_{\{1,4\}}-\overtr{LL}_{\{1\}} \overtr{VH_4}_{\{1,4\}}\right)\nn&\newline \times \left(x_{\{1,4\}} \overtr{LH_2}_{\{1,2,4\}} \overtr{LV}_{\{1,4\}}-\overtr{VH_2}_{\{1,2,4\}}\right) \left(x_{\{1,2,4\}} \overtr{LH_3}_{\{1,2,3,4\}} \overtr{LV}_{\{1,2,4\}}-\overtr{VH_3}_{\{1,2,3,4\}}\right)\nn
	&-\overtr{VH_4} \left(\overtr{LV}_{\{1\}} \overtr{LH_5}_{\{1,5\}}-\overtr{LL}_{\{1\}} \overtr{VH_5}_{\{1,5\}}\right) \left(x_{\{1,5\}} \overtr{LH_2}_{\{1,2,5\}} \overtr{LV}_{\{1,5\}}-\overtr{VH_2}_{\{1,2,5\}}\right)\nn&\newline\times \left(x_{\{1,2,5\}} \overtr{LH_3}_{\{1,2,3,5\}} \overtr{LV}_{\{1,2,5\}}-\overtr{VH_3}_{\{1,2,3,5\}}\right)\Bigg)+\text{permutations}\ \text{of}\ (2,3,4,5)\ed
\end{align}
\bibliographystyle{JHEP}
\bibliography{reference}

\end{document}